\documentclass[12pt,a4paper]{article}

\usepackage[T1]{fontenc}
\usepackage{amsmath,amssymb}
\usepackage{graphicx}
\usepackage{color}

\usepackage{ifpdf}
\ifpdf
  \usepackage[pdftex,
    pdftitle={Lyapunov exponent for inertial particles},
    pdfauthor={Peter Horvai},
    bookmarks,
    pdfstartview={Fit},
    unicode,
    a4paper]
  {hyperref}
\else
  \usepackage[dvips,
    pdftitle={Lyapunov exponent for inertial particles},
    pdfauthor={Peter Horvai},
    bookmarks,
    pdfstartview={Fit},
    unicode,
    a4paper]
  {hyperref}
\fi

\newcommand{\PLOTDIR}{./}
\newcommand{\FIGDIR}{./}

\newcommand{\dd}{{\mathrm d}}
\newcommand{\RR}{{\mathbb{R}}}
\newcommand{\CC}{{\mathbb{C}}}
\newcommand{\NN}{{\mathbb{N}}}
\newcommand{\HH}{{\mathbb{H}}}
\newcommand{\ul}[1]{\underline{\smash[b]{#1}}}

\newcommand{\mop}[1]{\mathop{\operatorname{#1}}}
\newcommand{\diag}{\mop{diag}\nolimits}
\newcommand{\tr}{\mop{tr}}
\newcommand{\Ito}{It\=o }

\title{Lyapunov exponent for inertial particles in the 2D Kraichnan
  model as a problem of Anderson localization with complex valued
  potential}
\author{P\'eter Horvai\thanks{
  Work done at Universit\`a di Roma ``La Sapienza'', Italy}
  \thanks{New address at University of Warwick, UK}
}

\begin{document}

\maketitle

\abstract{We exploit the analogy between dynamics of inertial particle
  pair separation in a random-in-time flow and the Anderson model of a
  quantum particle on the line in a spatially random real-valued
  potential.  Thereby we get an exact formula for the Lyapunov
  exponent of pair separation in a special case, and we are able to
  generalize the class of solvable models slightly, for potentials
  that are real up to a global complex multiplier.  A further important
  result for inertial particle behavior, supported by analytical
  computations in some cases and by numerics more generally, is that
  of the decay of the Lyapunov exponent with large Stokes number
  (quotient of particle relaxation and flow turn-over time-scales) as
  $\mathit{St}^{-2/3}$.}

\tableofcontents

\clearpage

\section{Introduction}

We propose in this paper to study evolution of inertial particle pair
separation in a smooth random flow.  The dynamics of pair separation,
through a very simple change of variables, is equivalent to the
Anderson localization problem \cite{MR0187859} of a quantum particle
in one-dimensional space with random potential, with the twist that
the potential is not necessarily a real-valued function.  This analogy
permits us to calculate explicitly, in certain situations, the
Lyapunov exponent of inertial particle pair separation.  In our
slightly generalized Anderson problem with complex potential, we find
some additional cases when the Lyapunov exponent may be computed
analytically.  The article treats the topic from the viewpoint of
inertial particles, but some results could also be interesting for the
Anderson localization problem.

Understanding the behavior of inertial particles, and more
specifically aerosols (tiny liquid or solid particles in a gas), has
several important applications.  For example rain formation inside
clouds depends on collisions of water droplets \cite{rainColl}, and a
better understanding of this process is needed for reliable weather
forecasts.  We may also mention internal combustion engines (oil
droplets, coal powder) where knowledge of particle behavior helps to
optimize design.  Since we will mostly study here the two dimensional
situation, let us also mention the problem of particles floating on
the free surface of a liquid, such as the surface of the ocean
\cite{MR1897722}.

The Lyapunov exponent of particle separation is one of the most
fundamental quantities describing relative motion of close-by
particles.  Finer details require knowledge of the distribution of
finite-time Lyapunov exponents, which would permit to predict such
things as collision rate of particles \cite{Collision}, to give an
example.  This paper is a first step in this direction.

Let us outline the plan of the paper.  First we give a presentation of
the smooth Kraichnan flow in which our particles shall evolve.  We
then describe dynamics of inertial particles and in particular of pair
separation, which we reduce to a two-dimensional problem.  After that
we treat the special case of genuinely 2D flows.  Later we exhibit
some additional solvable cases which generalize that of the classical
Anderson model.  We then come to results of numerical simulations that
sustain our findings and finally discuss inexactness of Piterbarg's
formula in \cite{MR1897722} as compared with numerics.

This paper also gives a review of the topic from the point of view of
inertial particles, so that results already known are rederived within
this framework and with our notations, thus serving possibly as
reference for further investigation in this direction.  Effort was
made to give detailed and clear calculations.

The author would like to thank Krzysztof Gaw\k{e}dzki for sharing his
ideas on the problem and pointing out the relationship with the
Anderson model.  Lot of credit goes to Alexander Fouxon, who
collaborated in this work, details of which are to appear in a common
paper \cite{2dlyapShort}; this collaboration was made possible thanks
to the European Network: ``Fluid mechanical stirring and mixing: the
Lagrangian approach''.  The author is grateful to Angelo Vulpiani who
offered the postdoc position and grant (funded by COFIN 2003:
``Statistical description of complex systems and systems with many
electrons'') while this work was prepared at the TNT group at the
Physics Department of the Universit\`a di Roma ``La Sapienza''.

\section{Smooth Kraichnan velocity field}

\subsection{The strain matrix}

We consider a smooth Kraichnan velocity field.  At small scale it may
be linearized so that velocity increments behave like
$$
  \vec U(\vec R + \vec r, t) - \vec U(\vec R, t)
=
  (\vec r . \vec \nabla_{\vec R}) \vec U (\vec R, t)
=
  r_j \partial_j U_i(\vec R, t) \vec e_i
$$
($\vec e_i$ are unit vectors of the chosen basis.  Also note the
convention that we use uppercase $R,U$ for absolute position and speed
while later we will use lowercase $r,v$ for the relative position and
speed of two particles.)  It is convenient to introduce the velocity
gradient matrix, also called strain matrix:
\begin{equation}
\label{eq:def-sigma}
  \sigma_{ij}(\vec R, t)
\equiv
  \partial_j U_i(\vec R, t)
\end{equation}
in order to write $\vec U(\vec R + \vec r, t) - \vec U(\vec R, t) =
\sigma(\vec R, t) \vec r$.

\subsection{Strain matrix correlation tensor}
\label{ssec:strain-correlation}

In the Kraichnan model the velocity field $U$ is a Gaussian,
white-in-time vector field, also called Brownian vector field.  We
suppose $\sigma(\vec R, t)$ to be a mean zero process (for all $\vec
R, t$), so that it is completely determined by its second moment
$\langle \sigma(\vec R, t) \otimes \sigma(\vec R', t') \rangle$, which
is proportional to $\delta (t-t')$, because of the white-in-time
hypothesis.

What will be important for us in tracking particle pair separation is
the statistics of the strain matrix along the trajectory of the
reference particle, as can be seen from \eqref{eq:IP-sep}.  Suppose
that the reference particle follows the trajectory $\vec R(t)$, which
we shall always suppose {\em continuous}.  Then define
\begin{equation}
\label{def:sigma-t}
  \sigma(t)
=
  \sigma(\vec R(t) , t)
\end{equation}

Since the matrix field $\sigma$ is centered Gaussian delta-correlated
in time, the same will hold for $\sigma(t)$.  Thus it is enough to
know the equal-time two-point function of $\sigma(t)$ to completely
characterize it.  In fact we have
\begin{align}
\notag
  \langle \sigma(t) \otimes \sigma(t') \rangle
&=
  \langle \sigma(\vec R(t), t) \otimes \sigma(\vec R(t'), t') \rangle
\\
\label{eq:sigma-t-2pt}
&=
  \langle \sigma(\vec R(t), t) \otimes \sigma(\vec R(t), t') \rangle
\end{align}
where the first equality follows from the definition
\eqref{def:sigma-t} of $\sigma(t)$, and the second equality holds
because the correlator on its left hand side being proportional to
$\delta(t-t')$, one may replace $\vec R(t')$ by $\vec R(t)$, since we
suppose $\vec R(t)$ continuous in $t$ and we suppose the two-point
correlator of the field $\sigma$ to be continuous in the spatial
variable.

Let us now introduce the equal-time equal-position correlation tensor
$C_{ik,jl}$ of the field $\sigma$, which may be defined through the
equality
\begin{equation}
\label{def:Cikjl}
  \langle \sigma_{ik}(\vec R, t) \sigma_{jl}(\vec R, t') \rangle
=
  \delta(t-t') \, C_{ik,jl}
\end{equation}
Since we suppose the statistics of the field $\sigma$ homogeneous in
space and time, $C_{ik,jl}$ is a constant independent of position and
time.  In Sect.~\ref{ssec:broken-sym} we point out that for the
velocity field $\vec U$ it is enough to have spatially homogeneous
increments, without $\vec U$ itself being spatially homogeneous, for
$\sigma$ to be homogeneous in the position variable.

Combining \eqref{eq:sigma-t-2pt} and \eqref{def:Cikjl} we have
\begin{equation}
\label{eq:sigma-t-C}
  \langle \sigma_{ik}(t) \sigma_{jl}(t') \rangle
=
  \delta(t-t') \, C_{ik,jl}
\end{equation}
In the setup of the Kraichnan model, where $\sigma(t)$ is Gaussian
white noise in time, it is often better to introduce and use
\begin{align}
\label{eq:def-dS}
  \dd S
&=
  \sigma(t) \dd t
\end{align}
and then we can talk about the covariance (better called {\em
covariance process} in the mathematical literature) of $\dd S$ and
write the more correct
$$
  \langle \dd S_{ik}, \dd S_{jl} \rangle
=
  C_{ik,jl} \,\dd t
$$

The covariance of $\sigma(t)$ may be derived from the velocity
two-point function $\langle U_i(\vec R, t) U_j(\vec R', t') \rangle$
as
\begin{align}
  \langle \sigma_{ik}(\vec R, t) \sigma_{jl}(\vec R, t') \rangle
&=
  \langle
    (\partial_k U_i(\vec R, t)) (\partial_l U_j(\vec R, t'))
  \rangle
\\
\label{eq:sigma-corr-nonhom}
&=
  \partial_{R_k} \partial_{R'_l}
  \langle
    U_i(\vec R, t) U_j(\vec R', t')
  \rangle |_{\vec R' = \vec R}
\end{align}

\subsection{Translation invariance}
\label{ssec:inhomogen}

Supposing spatial homogeneity of the velocity statistics, which may be
expressed as
$$
  \langle U_i(\vec R, t) U_j(\vec R', t') \rangle
=
  \delta(t-t') D_{ij}(\vec R - \vec R')
$$
for some tensor-valued function $D$, permits to further write
\eqref{eq:sigma-corr-nonhom} as
$$
  \langle \sigma_{ik}(\vec R, t) \sigma_{jl}(\vec R, t') \rangle
=
  - \delta(t-t') (\partial_k \partial_l D_{ij})(\vec 0)
$$
which in particular, upon substitution into \eqref{def:Cikjl}, leads
to the relationship
\begin{equation}
\label{eq:Cikjl-Dij}
  C_{ik,jl}
=
  -(\partial_k \partial_l D_{ij})(\vec 0)
\end{equation}
We thus arrive at the important conclusion that if the statistics of
the velocity field $\vec U$ is spatially homogeneous, then $C_{ik,jl}$
is symmetric under exchange of $k$ with $l$, or equivalently of $i$
with $j$, since the symmetry under simultaneous exchange of both pairs
is granted by the definition \eqref{def:Cikjl}.

Note however that one can imagine flows where, at least in some
bounded region of space, statistics of $\sigma$ {\em is} homogeneous
but that of the velocity field $\vec U$ is {\em not}.  For example a
uniformly expanding flow is like that.  To give examples of such a
flow, one can imagine either the surface flow of some liquid with
adequate up- and down-welling in some region, or the surface of a
bubble which grows and shrinks in time (but that is a curved surface,
and we do not pretend to treat that situation in this paper), or
perhaps a flat film whose boundary is attached to a growing and
shrinking ring.

\subsection{Isotropy}

Coming back to the case when the velocity field {\em is} spatially
homogeneous, if furthermore we suppose it isotropic and not breaking
parity symmetry (defined here as orthogonal transformations of
determinant $-1$), we have the second-order development for $D$ (the
velocity field is smooth!)
$$
  D_{ij}(\vec r)
\approx
  D_0 \delta_{ij} -
  D_1[(d+1-2\wp) \delta_{ij} r^2 + 2(\wp d  - 1) r_i r_j] +
  o(r^2)
$$
where $\wp$ is the compressibility degree \cite{MR1878800} of the flow
and $D_1$ is a dimensional constant of dimension $\mathsf{time}^{-1}$.
Hence, from \eqref{eq:Cikjl-Dij}
$$
  C_{ik,jl}
=
  -(\partial_k \partial_l D_{ij})(\vec 0)
=
  2 D_1  [(d+1-2\wp) \delta_{ij} \delta_{kl} +
          (\wp d  - 1) (\delta_{ik}\delta_{jl} +
                        \delta_{il}\delta_{jk})
         ]
$$

\subsection{Diagonalization, positivity}

As a side note, it is interesting to see that $C_{ik,jl}$ is indeed a
positive matrix, and if we wanted to generate $\sigma$ from
independent white noises, it would be useful also to be able to
diagonalize $C_{ik,jl}$.  We start with the latter, and criteria for
positivity will be obvious then.

Consider $C_{ik,jl}$ as the matrix of a linear transformation acting
on $d \times d$ matrices by
\begin{equation}
\label{eq:C-act}
  \sigma_{jl}
\mapsto
  \sigma_{ik}
=
  C_{ik,jl} \sigma_{jl}
\end{equation}
We identify three terms in $C_{ik,jl}$, namely $\delta_{ij}
\delta_{kl}$, $\delta_{ik}\delta_{jl}$ and $\delta_{il}\delta_{jk}$.
The first one is clearly the identity, which we shall conveniently
denote $\mop{Id}$.  The second acts as $M \mapsto (\tr M) 1_d$, (where
$\mop{1_d}$ is the $d \times d$ identity matrix, that is
$\diag_d(1,\ldots,1)$) and we shall call it $\mop{Tr}$.  The third acts
as transposition and we will call it $\mop{Tp}$.

From this it is clear that the three operators $\mop{Id}$, $\mop{Tr}$
and $\mop{Tp}$ commute mutually and thus may be diagonalized in a
common basis.  It is also straightforward to find this basis: use as
eigenspaces the space `$\operatorname{asym}$' of anti-symmetric
matrices, the space `$\operatorname{sym0}$' of symmetric traceless
matrices, and `$\operatorname{scal}$' of scalar matrices.  We have the
following eigenvalue structure:
$$
\begin{array}{l||c|c|c||c}
  & \mop{Id} & \mop{Tr} & \mop{Tp} & C_{ik,jl} \\ \hline\hline
  \operatorname{asym} & 1 & 0 & -1 & 2D_1 (d+2)(1-\wp) \\ \hline
  \operatorname{sym0} & 1 & 0 &  1 & 2D_1 [(d-2)(\wp+1)+2] \\ \hline
  \operatorname{scal} & 1 & d &  1 & 2D_1 \wp(d+2)(d-1) \\
\end{array}
$$

For $C_{ik,jl}$ to be positive on each eigenspace we get thus,
supposing $d \geq 2$ and $D_1>0$, the necessary and sufficient
condition $0 \leq \wp \leq 1$.

Since the correlation matrix $C_{ik,jl}$ is symmetric, it can be
diagonalized in a basis which is orthonormal with respect to the
scalar product induced by the identity matrix, which here means
$\delta_{ij} \delta_{kl}$ (see that this tensor is equal to 1 exactly
when the ordered pairs $(i,k)$ and $(j,l)$ coincide, and equal to 0
otherwise), and this scalar product is simply the sum of products of
corresponding matrix elements, formally given by $(\sigma,\sigma') =
\sigma_{ik} \sigma'_{ik}$.  In order to generate a (vector-valued)
Gaussian random variable with covariance matrix $C_{ik,jl}$, it is
enough to multiply each element (an eigenvector) of a given
orthonormal diagonalizing basis of $C_{ik,jl}$ by the square-root of
the corresponding eigenvalue and by an independent (for each basis
element) standard normal random variable.

In the case of $d=2$ dimensions, we can choose the following
orthonormal diagonalizing basis:
\begin{equation}
\label{tab:2d-diag}
\begin{array}{c||c|c||c}
  \operatorname{asym}
&
  \multicolumn{2}{|c||}{\operatorname{sym0}}
&
  \operatorname{scal}
\\
\hline
\hline
& & & \\[-1.8ex]
  \frac{1}{\sqrt{2}} \begin{pmatrix} 0 & 1 \\ -1 & 0 \end{pmatrix}
&
  \frac{1}{\sqrt{2}} \begin{pmatrix} 1 & 0 \\ 0 & -1 \end{pmatrix}
&
  \frac{1}{\sqrt{2}} \begin{pmatrix} 0 & 1 \\ 1 & 0 \end{pmatrix}
&
  \frac{1}{\sqrt{2}} \begin{pmatrix} 1 & 0 \\ 0 & 1 \end{pmatrix}
\end{array}
\end{equation}

In the case of $d=3$ dimensions, we can choose the following
orthonormal diagonalizing basis:
$$
\begin{array}{c||c|c||c}
  \operatorname{asym}
&
  \multicolumn{2}{|c||}{\operatorname{sym0}}
&
  \operatorname{scal}
\\
\hline
\hline
& & & \\[-1.8ex]
  \frac{1}{\sqrt{2}}
  \begin{pmatrix} 0 & 1 & 0 \\ -1 & 0 & 0 \\ 0 & 0 & 0 \end{pmatrix}
&
  \frac{1}{\sqrt{2}}
  \begin{pmatrix} 0 & 1 & 0 \\  1 & 0 & 0 \\ 0 & 0 & 0 \end{pmatrix}
&
  \frac{1}{\sqrt{2}}
  \begin{pmatrix} 0 & 0 & 1 \\ 0 & 0 & 0 \\  1 & 0 & 0 \end{pmatrix}
&
  \frac{1}{\sqrt{3}}
  \begin{pmatrix} 1 & 0 & 0 \\ 0 & 1 & 0 \\  0 & 0 & 1 \end{pmatrix}
\\[5ex]
  \frac{1}{\sqrt{2}}
  \begin{pmatrix} 0 & 0 & 1 \\ 0 & 0 & 0 \\ -1 & 0 & 0 \end{pmatrix}
&
  \frac{1}{\sqrt{2}}
  \begin{pmatrix} 0 & 0 & 0 \\ 0 & 0 & 1 \\ 0 &  1 & 0 \end{pmatrix}
&
  \frac{1}{\sqrt{2}}
  \begin{pmatrix} 1 & 0 & 0 \\ 0 & -1 & 0 \\  0 & 0 & 0 \end{pmatrix}
&
\\[5ex]
  \frac{1}{\sqrt{2}}
  \begin{pmatrix} 0 & 0 & 0 \\ 0 & 0 & 1 \\ 0 & -1 & 0 \end{pmatrix}
&
  \multicolumn{2}{c||}{
    \frac{1}{\sqrt{6}}
    \begin{pmatrix} 1 & 0 & 0 \\ 0 & 1 & 0 \\  0 & 0 & -2 \end{pmatrix}
  }
&
\end{array}
$$

\subsection{The 2D case with broken symmetries}
\label{ssec:broken-sym}

It is interesting to study more in detail the case of two-dimensional
flows where the velocity field's statistics' translation invariance or
parity invariance or both are not assumed.  We will however still
assume isotropy.

As mentioned in Sect.~\ref{ssec:inhomogen}, translation invariance is
broken if we don't have the symmetry of $C_{ik,jl}$ under exchange of
$k$ with $l$, or equivalently of $i$ with $j$.  This allows us to have
different coefficients for $\delta_{ik} \delta_{jl}$ and $\delta_{il}
\delta_{jk}$ in $C_{ik,jl}$.

Not supposing parity invariance allows for introduction of an
additional term in $C_{ik,jl}$, which can be written as $\epsilon_{ik}
\delta_{jl} + \epsilon_{jl} \delta_{ik}$ (recall that simultaneous
exchange of $i$ with $j$ and of $k$ with $l$ has to leave $C_{ik,jl}$
invariant).  Other expressions of the kind (one $\epsilon$ and one
$\delta$) give either a zero tensor or the same one (up to a possible
sign change).  In particular (see next paragraph for idea of simple
proof)
$$
  \epsilon_{ik} \delta_{jl} + \epsilon_{jl} \delta_{ik}
=
  \epsilon_{il} \delta_{jk} + \epsilon_{jk} \delta_{il}
$$
so that this term does not lead to breaking spatial homogeneity of the
velocity field.  Expressions with two $\epsilon$ can be expressed with
$\delta$s only.

The covariance matrix $C_{ik,jl}$ is completely characterized by its
action as defined in \eqref{eq:C-act}, and this action is linear.  In
particular it can be expressed as a matrix in the basis of $2 \times
2$ matrices of \eqref{tab:2d-diag}.  For convenience, we order the
basis elements as follows: the first two are those in
$\operatorname{sym0}$ , the third is of $\operatorname{asym}$ and the
fourth is that of $\operatorname{scal}$.  Then we have the following
actions:
$$
\begin{array}{c|c|c|c}
  \delta_{ij} \delta_{kl} &
  \delta_{ik} \delta_{jl} &
  \delta_{il} \delta_{jk} &
  \epsilon_{ik} \delta_{jl} + \epsilon_{jl} \delta_{ik}
\\[.8ex]
\hline
\hline
&&&\\[-1ex]
  \begin{pmatrix}
    1 & 0 & 0 & 0 \\
    0 & 1 & 0 & 0 \\
    0 & 0 & 1 & 0 \\
    0 & 0 & 0 & 1
  \end{pmatrix}
&
  \begin{pmatrix}
    0 & 0 & 0 & 0 \\
    0 & 0 & 0 & 0 \\
    0 & 0 & 0 & 0 \\
    0 & 0 & 0 & 2
  \end{pmatrix}
&
  \begin{pmatrix}
    1 & 0 &  0 & 0 \\
    0 & 1 &  0 & 0 \\
    0 & 0 & -1 & 0 \\
    0 & 0 &  0 & 1
  \end{pmatrix}
&
  \begin{pmatrix}
    0 & 0 & 0 & 0 \\
    0 & 0 & 0 & 0 \\
    0 & 0 & 0 & 2 \\
    0 & 0 & 2 & 0
  \end{pmatrix}
\end{array}
$$
This representation is convenient also for finding identities between
different expressions.  For example we have the action
$$
  \epsilon_{ik} \epsilon_{jl}
\quad:\quad
  \begin{pmatrix}
    0 & 0 & 0 & 0 \\
    0 & 0 & 0 & 0 \\
    0 & 0 & 2 & 0 \\
    0 & 0 & 0 & 0
  \end{pmatrix}
$$
from which we see that
\begin{equation}
\label{eq:epsilon-delta}
  \epsilon_{ik} \epsilon_{jl}
=
  \delta_{ij} \delta_{kl} - \delta_{il} \delta_{jk}
\end{equation}

Our next task is to find out when is $C_{ik,jl}$, constructed from the
above four tensors, a positive matrix.  For this, we see first that in
general (for generic coefficients $a,b,c,d \in \RR$) we have the
action
$$
  a \delta_{ij} \delta_{kl} +
  b \delta_{ik} \delta_{jl} +
  c \delta_{il} \delta_{jk} +
  d (\epsilon_{ik} \delta_{jl} + \epsilon_{jl} \delta_{ik})
\ :\ 
  \begin{pmatrix}
    a+c &  0  &  0  &    0   \\
     0  & a+c &  0  &    0   \\
     0  &  0  & a-c &   2d   \\
     0  &  0  &  2d & a+2b+c
  \end{pmatrix}
$$
The necessary and sufficient condition for $C_{ik,jl}$ to be positive
is that all eigenvalues of the above matrix are positive.  These
eigenvalues are readily found to be
$$
  a+c \quad\text{(twice)}, \qquad a+b \pm \sqrt{(b+c)^2+4d^2}
$$
The term under the square-root is always positive, and since the
square-root itself is defined to be positive, it is enough to retain
the two criteria
\begin{gather}
\label{eq:pos1}
  a+c
\geq
  0
\\
\label{eq:pos2}
  a+b - \sqrt{(b+c)^2+4d^2}
\geq
  0
\end{gather}
as necessary and sufficient for $C_{ik,jl}$ to be positive.

\section{Inertial particles in the linearized flow}

\subsection{Basic equation}
\label{ssec:IP-basic-eq}

Let us turn to the inertial particles.  An inertial particle is one
whose velocity does not necessarily coincide with the speed of the
fluid flow surrounding the particle.  Instead its velocity relaxes to
the latter by viscous friction, at an exponential rate whose time
coefficient is the Stokes time, which we shall denote $\tau$.

It is not completely clear what equation to use to describe inertial
particles.  The work of Maxey and Riley \cite{MaxRi} is often cited
and ad hoc simplifications of their formula are made.  Also the fact
that describing particle separation is a different matter from
describing single-particle motion is mostly neglected.  Here we shall
admit that for very heavy particles (the case of aerosols) the
following equation is a good approximation both for particle movement
and for extracting from it the equation on particle separation.
Formally, we thus take a system where inertial particle evolution is
described by the very simple equation:
\begin{equation}
\label{eq:inertial-particles}
  \frac{\dd \vec R}{\dd t}
=
  \vec V
,\qquad
  \frac{\dd \vec V}{\dd t}
=
  -\frac{1}{\tau} (\vec V - \vec U(\vec R, t))
\end{equation}
Remark that, in the case of the Kraichnan model,
\eqref{eq:inertial-particles} is in fact a stochastic differential
equation (an SDE; see Sect.~\ref{ssec:pass-to-SDE} for a justification
of why \eqref{eq:inertial-particles} is an SDE: the idea is
illustrated on \eqref{eq:IP-sep}), so in principle we should also
specify what interpretation (\Ito, Stratonovich, etc.) we take, but
arguments of Sect.~\ref{ssec:conv-indiff} apply here too to show that
the particular interpretation convention doesn't matter.

Eq.~\eqref{eq:inertial-particles} gives for the infinitesimal
separation $\vec r$ and relative velocity $\vec v$ of two
(infinitesimally separated) particles:
\begin{equation}
\label{eq:IP-sep}
  \frac{\dd \vec r}{\dd t}
=
  \vec v
,\qquad
  \frac{\dd \vec v}{\dd t}
=
  -\frac{1}{\tau} (\vec v - \sigma(t) \vec r)
\end{equation}
where $\sigma(t)$ is as defined in \eqref{def:sigma-t}, and in
particular depends on the trajectory $\vec R(t)$ of the reference
particle, so that in principle \eqref{eq:IP-sep} is not autonomous and
should be coupled to \eqref{eq:inertial-particles}.  But for the
purpose of calculating {\em statistical averages} of the particle
separation it is quite enough to know just the {\em distribution} of
the process $\sigma(t)$.  As was already discussed in
Sect.~\ref{ssec:strain-correlation}, in the case of a Kraichnan
velocity field the process $\sigma(t)$ is centered Gaussian white
noise with two-point function given by \eqref{eq:sigma-t-C}, and this
is all that we need to know.  Once again, for the Kraichnan velocity
field, \eqref{eq:IP-sep} is an SDE, and again interpretation
convention doesn't matter, as developed in
Sect.~\ref{ssec:pass-to-SDE} and \ref{ssec:conv-indiff}.

\subsection{Anderson localization form}
\label{ssec:Anderson-form}

The two first-order differential equations of \eqref{eq:IP-sep} can be
combined to one second-order on $\vec r$
$$
  \frac{\dd^2 \vec r}{\dd t^2}
=
  -\frac{1}{\tau} \frac{\dd \vec r}{\dd t}
  +\frac{\sigma}{\tau} \vec r
$$
or, introducing $\vec \psi(t) = e^{t/2\tau} \, \vec r$ to eliminate
the first-order term:
$$
  \frac{\dd^2 \vec \psi}{\dd t^2}
=
  \left(\frac{1}{4\tau^2} + \frac{\sigma}{\tau}\right) \vec \psi
$$
Introduce
\begin{equation}
\label{def:E,V}
  E = -1/4\tau^2 \ \ \text{(energy)},
\qquad
  V = \sigma/\tau \ \ \text{(matrix white-noise potential)},
\end{equation}
consider $t$ to be position rather than time, and we have the
Schr\"{o}dinger eigenvalue equation associated to the Anderson
localization problem (though the wave function $\vec\psi$ takes values
in $\RR^d$ not simply in $\RR$, and the potential $V$ is accordingly
matrix valued, not simply real):
\begin{equation}
\label{eq:Anderson-multi}
  -\frac{\dd^2 \vec \psi}{\dd t^2} + V \vec \psi
=
  E \vec \psi
\end{equation}
This analogy between the system \eqref{eq:IP-sep} and the Anderson
model is hinted at (and used) in \cite{PRL2506021}.

In the one-dimensional case (our $d=1$), the Lyapunov exponent
associated to the growth-rate of $|\vec\psi|$ (more exactly
of $\sqrt{\smash[b]{\vec\psi^2+\vec{\psi'}{}^2}}$) is known (cf.\
Sect.~\ref{ssec:sol-case-1}), but generally not in the
multi-dimensional case.

\subsection{Analytical form}
\label{ssec:Analytic-form}

The equation of evolution may also be written for matrices
$\mathcal{R}$ and $\mathcal{V}$ instead of particle separations $\vec
r$ and $\vec v$:
$$
  \frac{\dd\mathcal{R}}{\dd t}
=
  \mathcal{V}
,\qquad
  \frac{\dd\mathcal{V}}{\dd t}
=
  -\frac{1}{\tau} (\mathcal{V} - \sigma(t) \mathcal{R})
$$
where every corresponding column of $\mathcal{R}$ and $\mathcal{V}$
correspond to a couple $\vec r$ and $\vec v$, evolving in the same
velocity field but independently of the other columns.

It is straightforward to see that the matrix
$\tilde{\mathcal{Z}} = \mathcal{V} \mathcal{R}^{-1}$ verifies the
first-order equation
$$
  \frac{\dd\tilde{\mathcal{Z}}}{\dd t}
=
  - \tilde{\mathcal{Z}}^2
  - \frac{1}{\tau} \tilde{\mathcal{Z}} + \frac{\sigma}{\tau}
$$
since for any matrix-valued function $\dd\mathcal{R}^{-1} = -
\mathcal{R}^{-1} (\dd\mathcal{R}) \mathcal{R}^{-1}$.  We may also
introduce $\mathcal{Z} = \tilde{\mathcal{Z}}+1/2\tau$ and write
$$
  \frac{\dd\mathcal{Z}}{\dd t}
=
  - \mathcal{Z}^2 + \frac{1}{4\tau^2} + \frac{\sigma}{\tau}
$$

\subsection{Passing to an SDE}
\label{ssec:pass-to-SDE}

Since $\sigma(t)$ is a finite-dimensional Gaussian process, it may be
represented as linear combinations of a finite number (at most $d^2$)
of independent white noises.  Denoting the latter by $\ul{\dd w}$
(with $\langle \dd w_i , \dd w_j \rangle = \delta_{ij} \dd t$) clearly
we may write, for some (not necessarily square) matrix $B(\vec r)$
$$
  \tau^{-1} \sigma(t) \vec r
=
  B(\vec r) \, \ul{\dd w} / \dd t
$$
or equivalently
\begin{equation}
\label{eq:def-B}
  \tau^{-1} \dd S \vec r
=
  B(\vec r) \, \ul{\dd w}
\end{equation}

With this in mind, we may write the block-matrix form equation
\begin{equation}
\label{eq:IPSDE}
  \dd \begin{pmatrix} \vec r \\ \vec v \end{pmatrix}
=
   \begin{pmatrix} \vec v \\ -\frac{1}{\tau} \vec v \end{pmatrix} \dd t
  +\begin{pmatrix} 0 \\ B(\vec r) \end{pmatrix} \ul{\dd w}
\end{equation}
The merit of this formulation is that we immediately see that we have
a stochastic differential equation.

\subsection{Whatever convention (\Ito etc.)}
\label{ssec:conv-indiff}

Furthermore it doesn't matter what convention (\Ito, etc.) we use to
interpret \eqref{eq:IPSDE}, as we shall now show.  Indeed, write the
generic equation
\begin{equation}
\label{eq:genSDE}
  \dd \vec X
=
  \vec A \dd t + M \ul{\dd w}
\end{equation}
where $\vec A$ and $B$ depend on $\vec X$.  Then from passing from \Ito
to Stratonovich convention, we have the additional term $\frac{1}{2}
\langle (((M \ul{\dd w}) . \vec \nabla) M) \ul{\dd w} \rangle$, easier
to develop in coordinate notation:
$$
  \frac{1}{2}
  \langle M_{ij} \dd w_j (\partial_i M_{kl}) \dd w_l \rangle 
=
  \frac{1}{2}
  (M_{ij} \partial_i M_{kl}) \langle \dd w_j \dd w_l \rangle
=
  \frac{1}{2} M_{ij} \partial_i M_{kj} \, \dd t
$$
since $\langle \dd w_j \dd w_l \rangle = \delta_{jl} \dd t$.  It is
easy to check that in our case the above is 0 since $M_{ij}
\partial_i$ will only involve derivatives with respect to $\vec v$
whereas $M_{kj}$ depends only on $\vec r$.

The same reasoning as above applies to single-particle motion
described by \eqref{eq:inertial-particles}.  It is then interesting to
note that the $\tau \to 0$ limit may be taken and we thus get a {\em
unique} convention both for the passive tracer particle and the
separation of such particles, which can be seen to be the \Ito
convention (cf.\ Appendix \ref{app:tau-to-0}).  In other words, in a
Kraichnan flow a passive tracer that is the $\tau\to0$ limit of an
inertial particle is solution of the advection SDE with \Ito
convention.

\subsection{The Markov process generated by the SDE}

Now let us look at the Markov process generated by an SDE of form
\eqref{eq:genSDE}.  We may write the probability density function of
$\vec X$ as $P(\vec x) = \langle \delta(\vec X -\vec x) \rangle$, and
use the \Ito formula
$$
  \dd f(\vec X)
=
   \dd X_i (\partial_i f) (\vec X)
  +\frac{1}{2}
     \langle \dd X_i \dd X_j \rangle
     (\partial_i \partial_j) f(\vec X)
$$
to obtain
\begin{align*}
  \dd P(\vec x)
&=
  \dd \langle \delta(\vec X - \vec x) \rangle
=
  \langle \dd \delta(\vec X - \vec x) \rangle
\\
&=
  -A_i \partial_{x_i} \langle \delta(\vec X - \vec x) \rangle
  +\frac{1}{2} \langle M_{ik} \dd w_k M_{jl} \dd w_l \rangle
     \partial_{x_i} \partial_{x_j}
       \langle \delta(\vec X - \vec x) \rangle
\\
&=
  \left[
    -A_i (\partial_i P) (\vec x)
    +\frac{1}{2} M_{ik} M_{jk} (\partial_i \partial_j P) (\vec x)
  \right]
  \dd t
\end{align*}
since $\langle \dd w_k \dd w_l \rangle = \delta_{kl} \dd t$.  Thus for
the Markov process generated by the SDE \eqref{eq:genSDE} it is not
$M$ and $w$ itself that are important, but only $M_{ik} M_{jk}$, in
other words $MM^T$, so that \eqref{eq:genSDE} may be replaced by the
effective equation
\begin{equation}
\label{eq:eff-genSDE}
  \dd \vec X
=
  \vec A \dd t + \tilde M \ul{\dd \tilde w}
\end{equation}
for some driving noise $\ul{\dd\tilde w}$, not necessarily of the same
dimension as $\ul{\dd w}$, with $\langle \dd\tilde w_i \dd\tilde w_j
\rangle = \delta_{ij} \dd t$.  Note how this fixes only the number of
rows of $\tilde M$ (to be the same as the dimension of $\vec X$), but
not the number of its columns.

\subsection{Reducing the dimension of the driving noise}
\label{ssec:red-dim}

Applying the general considerations of the preceding section to our
case, with the particular form \eqref{eq:IPSDE} of \eqref{eq:genSDE},
i.e.\ having
$$
  \vec X
=
  \begin{pmatrix} \vec r \\ \vec v \end{pmatrix}
\qquad\qquad
  \vec A
=
  \begin{pmatrix} \vec v \\ -\frac{1}{\tau} \vec v \end{pmatrix}
\qquad\qquad
  M
=
  \begin{pmatrix} 0 \\ B(\vec r) \end{pmatrix}
$$
we have
$$
  M M^T
=
  \begin{pmatrix} 0 & 0 \\ 0 & B(\vec r) B^T(\vec r) \end{pmatrix}
$$
and using formula \eqref{eq:def-B} for $B(\vec r)$ we get
\begin{align}
\label{eq:BBT-stoch}
  (B B^T)_{ij} (\vec r)
&=
  B_{ik}(\vec r) B_{jk}(\vec r)
=
  \langle B_{ik} \dd w_k B_{jl} \dd w_l \rangle / \dd t
=
  \langle \dd S_{ik} r_k \dd S_{jl} r_l \rangle / \dd t
\\
\label{eq:Crr}
&=
  \tau^{-2} C_{ik,jl} r_k r_l
=
  \frac{2 D_1}{\tau^2}
    [(d+1-2\wp) r^2 \delta_{ij} + 2(\wp d - 1) r_i r_j]
\end{align}
We may choose any other $\tilde{B}(\vec r)$ {\em with real
coefficients} giving the same result, i.e.\ such that:
\begin{equation}
\label{eq:B-Btilde}
  \tilde B_{ik}(\vec r) \tilde B_{jk}(\vec r)
=
  B_{ik}(\vec r) B_{jk}(\vec r)
=
  \tau^{-2} C_{ik,jl} r_k r_l
\end{equation}
and use it in \eqref{eq:eff-genSDE} by posing
$$
  \tilde M
=
  \begin{pmatrix} 0 \\ \tilde B(\vec r) \end{pmatrix}
$$

Notice that for any orthogonal basis formed of $\vec r$ and $d-1$
other vectors of length $r$, whose matrix we shall denote $R$ (i.e.\
the columns of $R$ are the vectors of the basis), we have
\begin{align*}
  B B^T
&=
  \frac{2 D_1}{\tau^2}
    [(d+1-2\wp) R R^T + 2(\wp d - 1) (R \vec e_1) (R \vec e_1)^T]
\intertext{since $R R^T = r^2 \diag_d(1,\ldots,1)$ and $\vec r = R \vec
  e_1$; then note $\vec e_1 \vec e_1^T = \diag_d(1,0,\ldots,0)$ and get}
  B B^T
&=
  \frac{2 D_1}{\tau^2}
    R \diag_d((2\wp+1)(d-1),d+1-2\wp, \ldots, d+1-2\wp) R^T
\end{align*}
(note $d+1-2\wp + 2(\wp d - 1) = (2\wp+1)(d-1)$) so that in particular
we may take
\begin{equation}
\label{eq:Btilde-Rbeta}
  \tilde{B}
=
  R \diag_d(\sqrt{\beta_L}, \sqrt{\beta_N}, \ldots, \sqrt{\beta_N})
\end{equation}
where we have introduced
\begin{equation}
\label{def:betas}
  \beta_L
=
  2 \tau^{-2} D_1 (2\wp+1) (d-1)
\qquad\qquad
  \beta_N
=
  2 \tau^{-2} D_1 (d+1-2\wp)
\end{equation}

The effective SDE we obtain for \eqref{eq:IPSDE} is driven by a
$d$-dimensional white-noise $\ul{\dd\tilde w}$ instead of the
$d^2$-dimensional white-noise $\ul{\dd w}$, and is written
\begin{equation}
\label{eq:SDE-red1}
  \dd \begin{pmatrix} \vec r \\ \vec v \end{pmatrix}
=
   \begin{pmatrix} \vec v \\ -\frac{1}{\tau} \vec v \end{pmatrix}
     \dd t 
  +\begin{pmatrix} 0 \\ \tilde{B}(\vec r) \end{pmatrix}
     \ul{\dd\tilde w}
\end{equation}

\subsection{An even stronger reduction}

Due to symmetries of the problem, a closed evolution equation may be
written on the magnitudes and relative angle of particle separation
$\vec r$ and relative speed $\vec v$.  Let us begin by introducing the
quantities
\begin{equation}
\label{def:XYZ}
  X = \vec v . \vec r
\qquad
  Y = \vec r . \vec r
\qquad
  Z = \vec v . \vec v
\end{equation}
These verify the evolution equations ({\em note that throughout we use
\Ito convention to interpret the SDEs below})
\begin{align*}
  \dd X
&=
  \dd \vec v . \vec r + \vec v . \dd \vec r
=
  [-\frac{1}{\tau} \vec v . \vec r +  \vec v . \vec v] \dd t +
  \frac{1}{\tau} \vec r . (\dd S \vec r)
\\
  \dd Y
&=
  2 \vec r . \dd \vec r
=
  2 \vec r . \vec v \dd t
\\
  \dd Z
&=
  2 \vec v . \dd \vec v + \langle \dd \vec v_i, \dd \vec v_i \rangle
=
  [-\frac{2}{\tau} \vec v . \vec v
   + \frac{2D_1}{\tau^2}(d+2)(d-1) \vec r . \vec r
  ] \dd t
  + \frac{2}{\tau} \vec v . (\dd S \vec r)
\end{align*}
where in the last line we used
\begin{align*}
  \langle \dd v_i, \dd v_i \rangle
&=
  \langle \dd S_{ik} r_k, \dd S_{il} r_l \rangle
=
  \langle \dd S_{ik}, \dd S_{il} \rangle r_k r_l
\\
&=
  C_{ik,il} \, \dd t \, r_k r_l
=
  2 D_1 [(d+1-2\wp)d + 2(\wp d - 1)] \, (\vec r . \vec r) \, \dd t
\\
&=
  2 D_1 (d+2) (d-1) \, Y \, \dd t
\end{align*}
It is convenient to introduce the new noises $\dd\eta_1$, $\dd\eta_2$
defined by $\dd\eta_1 = \vec r . (\dd S \vec r)$ and $\dd\eta_2 = \vec
v . (\dd S \vec r)$.  We only need to know their correlations:
\begin{align*}
  \langle \dd \eta_1, \dd \eta_1 \rangle
&=
  \langle \dd S_{ik} r_i r_k, \dd S_{jl} r_j r_l \rangle
=
  \langle \dd S_{ik}, \dd S_{jl} \rangle r_i r_k r_j r_l
\\
&=
  C_{ik,jl} \, \dd t \, r_i r_k r_j r_l
=
  2 D_1 (2\wp + 1)(d-1) \, (\vec r . \vec r)^2 \, \dd t
\end{align*}
In the same manner
\begin{align*}
  \langle \dd \eta_2, \dd \eta_2 \rangle
&=
  \langle \dd S_{ik} v_i r_k, \dd S_{jl} v_j r_l \rangle
=
  C_{ik,jl} \, \dd t \, v_i r_k v_j r_l
\\
&=
  2 D_1 [ (d+1-2\wp)  (\vec r . \vec r)(\vec v . \vec v)
         +2(\wp d - 1) (\vec v . \vec r)^2
        ] \, \dd t
\end{align*}
and
\begin{align*}
  \langle \dd \eta_1, \dd \eta_2 \rangle
&=
  \langle \dd S_{ik} r_i r_k, \dd S_{jl} v_j r_l \rangle
=
  C_{ik,jl} \, \dd t \, r_i r_k v_j r_l
\\
&=
  2 D_1 (2\wp+1)(d-1) \, (\vec v . \vec r) (\vec r . \vec r) \, \dd t
\end{align*}
To sum it up
\begin{align*}
  \dd X
&=
  [-\frac{1}{\tau} X + Z] \dd t +
  \frac{1}{\tau} \dd \eta_1
\\
  \dd Y
&=
  2 X \dd t
\\
  \dd Z
&=
  [-\frac{2}{\tau} Z + \frac{2D_1}{\tau^2}(d+2)(d-1) Y] \dd t
  + \frac{2}{\tau} \dd \eta_2
\end{align*}
with
\begin{align*}
  \langle \dd \eta_1, \dd \eta_1 \rangle
&=
  2 D_1 (2\wp + 1)(d-1) \, Y^2 \, \dd t
\\
  \langle \dd \eta_2, \dd \eta_2 \rangle
&=
  2 D_1 [(d+1-2\wp) YZ + 2(\wp d - 1) X^2] \, \dd t
\\
  \langle \dd \eta_1, \dd \eta_2 \rangle
&=
  2 D_1 (2\wp+1)(d-1) \, XY \, \dd t
\end{align*}

We can now go on by introducing
\begin{equation}
\label{def:AB} 
A = X/Y
\qquad\qquad\qquad
 B = Z/Y
\end{equation}
Then
\begin{align*}
  \dd A
&=
  \dd \frac{X}{Y}
=
  \frac{1}{Y} \dd X - \frac{X}{Y^2} \dd Y
=
  (-\frac{1}{\tau} A - 2 A^2 + B) \dd t + \frac{1}{\tau} \dd \gamma_1
\\
  \dd B
&=
  \dd \frac{Z}{Y}
=
  \frac{1}{Y} \dd Z - \frac{Z}{Y^2} \dd Y
=
  [-\frac{2}{\tau} B - 2AB + \frac{2D_1}{\tau^2} (d+2)(d-1)] \dd t
  + \frac{2}{\tau} \dd \gamma_2
\end{align*}
with $\dd \gamma_1 = Y^{-1} \dd \eta_1$ and $\dd \gamma_2 = Y^{-1}
\dd \gamma_2$, so that
\begin{align*}
  \langle \dd \gamma_1, \dd \gamma_1 \rangle
&=
  2 D_1 (2\wp + 1)(d-1) \, \dd t
\\
  \langle \dd \gamma_2, \dd \gamma_2 \rangle
&=
  2 D_1 [(d+1-2\wp) B + 2(\wp d - 1) A^2] \, \dd t
\\
  \langle \dd \gamma_1, \dd \gamma_2 \rangle
&=
  2 D_1 (2\wp+1)(d-1) \, A \, \dd t
\end{align*}

Introduce $C=B-A^2$.  Then
$$
  \dd C
=
  \dd B - 2 A \dd A - \langle \dd A, \dd A \rangle
=
  [-\frac{2}{\tau} C - 4AC + \frac{2D_1}{\tau^2} (d-1)(d+1-2\wp)] \,
  \dd t
  + \frac{2}{\tau} \dd \gamma_3
$$
where we have introduced $\dd \gamma_3 = \dd \gamma_2 - A \dd
\gamma_1$.  We have
\begin{align*}
  \langle \dd \gamma_3, \dd \gamma_3 \rangle
&=
       \langle \dd\gamma_2, \dd\gamma_2 \rangle
  +A^2 \langle \dd\gamma_1, \dd\gamma_1 \rangle
  -2A  \langle \dd\gamma_1, \dd\gamma_2 \rangle
\\
&=
  2 D_1 (d+1-2\wp) \, C \, \dd t
\\
  \langle \dd \gamma_1, \dd \gamma_3 \rangle
&=
  0
\end{align*}
Finally introducing $F=C^{1/2}$ we arrive at
\begin{align*}
  \dd F
&=
  \dd C^{1/2}
=
   \frac{1}{2} C^{-1/2} \dd C
  -\frac{1}{8} C^{-3/2} \langle \dd C, \dd C \rangle
\\
&=
  [-(\frac{1}{\tau} + 2A) F
   + \frac{D_1}{\tau^2} (d-2)(d+1-2\wp) F^{-1}] \, \dd t
  +\frac{1}{\tau} \dd \gamma_4
\end{align*}
with $\dd\gamma_4 = C^{-1/2}\dd\gamma_3$ so that
$$
  \langle \dd\gamma_4, \dd\gamma_4 \rangle
=
  2 D_1 (d+1-2\wp) \, \dd t
$$

At this point, let us introduce $z=A+iF+1/(2\tau)$ (so the real part
of $z$ is the (signed) component of $\vec v$ along $\vec r$, divided
by $r$ and then shifted by $1/(2\tau)$, while the imaginary part of
$z$ is the norm of the component of $\vec v$ perpendicular to $\vec
r$, divided by $r$) and $\dd\gamma = \dd\gamma_1 + i \dd\gamma_4$ to
have
$$
  \dd z
=
  [  \frac{1}{4\tau^2} - z^2
   + i \frac{D_1}{\tau^2} (d-2)(d+1-2\wp)(\Im z)^{-1}] \, \dd t
  + \frac{1}{\tau} \dd\gamma
$$
with (since $\langle \dd\gamma_1, \dd\gamma_4 \rangle = 0$)
$$
  \langle \dd\gamma, \dd\gamma \rangle
=
   \langle \dd\gamma_1, \dd\gamma_1 \rangle
  -\langle \dd\gamma_4, \dd\gamma_4 \rangle
=
  4D_1(\wp d - 1) \, \dd t
$$
Or, using definitions from \eqref{def:E,V} and \eqref{def:betas}, we
can write
\begin{equation}
\label{eq:SDE-red2}
  \dd z
=
  [- E - z^2
   + i \frac{d-2}{2} \beta_N (\Im z)^{-1}] \, \dd t
  + \dd\tilde\gamma
\end{equation}
with
\begin{equation}
\label{def:gammatilde}
  \dd\tilde\gamma = \dd\tilde\gamma_L + i \,\dd\tilde\gamma_N
\end{equation}
where $\dd\tilde\gamma_L$ and $\dd\tilde\gamma_N$ are real white
noises with correlation
\begin{equation}
\label{eq:corr-gammatilde}
  \langle \dd\tilde\gamma_L, \dd\tilde\gamma_L \rangle
=
  \beta_L \,\dd t
\qquad
  \langle \dd\tilde\gamma_N, \dd\tilde\gamma_N \rangle
=
  \beta_N \,\dd t
\qquad
  \langle \dd\tilde\gamma_L, \dd\tilde\gamma_N \rangle
=
  0
\end{equation}
so that in particular $\langle \dd \tilde \gamma, \dd \tilde \gamma
\rangle = (\beta_L - \beta_N) \,\dd t$.  We emphasize that $\gamma$
and $\tilde\gamma$ are Brownian motions {\em not} on the real line but
on the complex plane, each equivalent to a two-dimensional real
Brownian motion.  Formula \eqref{eq:SDE-red2} (with other notations)
appears in \cite{InAgg}.

We then see that the case $d=2$ is special since then the equation on
$z$ is analytical.  For $d > 2$ the additional drift comes from the
fact that we reduced the diffusion of $\vec v$ perpendicularly to
$\vec r$ to its radial part, exploiting rotational symmetry around the
direction of $\vec r$.  Then we have the well known inverse-radius
drift term of the radial Laplacian.

Also the case $(d+1-2\wp)=0$ is special, but as we will see in the
next subsection, it is a much more particular case, since $\vec v$
tends to align with $\vec r$ so that we recover the one-dimensional
situation which is explicitly solvable.

Since we also have (cf.\ Eqs.\ \eqref{def:XYZ}, \eqref{def:AB})
$$
  \dd \log |\vec r|
=
  \frac{1}{2Y} \, \dd Y
=
  A \, \dd t
=
  \Big( \Re z - \frac{1}{2\tau} \Big) \, \dd t
$$
we have the expression for the Lyapunov exponent of particle
separations
\begin{equation}
\label{eq:lambda-z}
  \lambda
= 
  \left\langle \frac{\dd \log |\vec r|}{\dd t} \right\rangle
=
  \langle \Re z \rangle - \frac{1}{2\tau}
\end{equation}

\subsection{A solvable case}
\label{ssec:sol-case-1}

The ``physical'' values for $\beta_N/\beta_L$ are those that
correspond to $0 \leq \wp \leq 1$, i.e.\ $\frac{1}{3} \leq
\beta_N/\beta_L \leq 1+\frac{2}{d-1}$, in the sense that only these
correspond to the evolution of the separation of inertial particles in
a linearized Kraichnan flow.  However, the reduced SDE
\eqref{eq:SDE-red1} or \eqref{eq:SDE-red2} is meaningful as long as
$\beta_L, \beta_N \geq 0$ (cf.\ respectively \eqref{eq:Btilde-Rbeta}
and \eqref{eq:corr-gammatilde}).

The case for which the Lyapunov exponent is known is when $\beta_N =
0$, and additionally the initial condition is such that $\vec v
\parallel \vec r$, as then \eqref{eq:SDE-red2} has real initial $z$
and both the imaginary term in the drift and in the diffusion (cf.\
\eqref{def:gammatilde}, \eqref{eq:corr-gammatilde}) disappear.  We
thus recover the classical Anderson localization problem: introduce
$\psi = e^{t/2\tau} r$ (here $r\in\RR$ is the signed coordinate of
$\vec r$ along the line defined by the initial condition $\vec r(0)$)
verifying
$$
  -\frac{\dd^2\psi}{\dd t^2} + \frac{\dd\tilde\gamma}{\dd t}\psi
=
  E \psi
$$
then $z = \psi'/\psi$ verifies \eqref{eq:SDE-red2}, so $\langle z
\rangle$ is the growth rate of $\ln|\psi|$ which we know from the
Anderson problem.  Hence, recalling the simple relationship
\eqref{eq:lambda-z} between $\langle z \rangle$ and $\lambda$, the
latter may be expressed in terms of the Airy functions of first and
second kind, $Ai$ and $Bi$ respectively, and their derivatives $Ai'$
and $Bi'$, as
\begin{equation}
\label{eq:lambda-AiBi}
  \lambda
=
  \frac{1}{2\tau}
  \left[
    -1 + c^{-1/2} \frac{Ai'(c)Ai(c) + Bi'(c)Bi(c)}{Ai^2(c) + Bi^2(c)}
  \right]
\end{equation}
where $c = 1/[4\tau^2(\beta_L/2)^{2/3}]$.

Later on we need to examine if the $\vec v \parallel \vec r$ initial
condition is really necessary for the above formula to hold.
Numerical simulations of \eqref{eq:IP-sep} (in fact of
\eqref{eq:psep-2D-red}, which is equivalent {\em even in $d\neq2$
dimensions}, since \eqref{eq:IP-sep} reduces to \eqref{eq:SDE-red2},
which for $\beta_N=0$ is identical to \eqref{eq:z-evol}, and the
latter is derived from \eqref{eq:psep-2D-red} and is sufficient to
give the corresponding Lyapunov exponent) show that (at least for the
examined values of parameters), this condition is irrelevant.

It is tempting then to think that this irrelevance of the initial
alignment of $\vec v$ with $\vec r$ is due to the asymptotic
``collapse'' of $\vec v$ onto the direction of $\vec r$ at long times.
In other words, that $z$, even if initially it has an imaginary part,
can be considered to be real at large times, because it would be
``attracted'' to the real axis by the dynamics of its evolution.
However this is {\em not} the case, as evidenced by numerical
simulations, since $|\Re z|$ has a finite limit when $t \to \infty$,
which in fact we can also calculate.  The more general treatment of
Sect.~\ref{sec:other-sol} will shed light on this and justify the
expression \eqref{eq:lambda-AiBi} even for non-aligned initial $\vec
v$ and $\vec r$.

\section{The 2D case}
\label{sec:2D}

\subsection{Passing to the complex notation in 2D}

Recalling Sect.~\ref{ssec:red-dim} and in particular the requirement
\eqref{eq:B-Btilde} (see also the last equality of \eqref{eq:Crr}), we
can choose for $\tilde B(\vec r)$, in the 2 dimensional case, the
particular form
$$
  \tilde{B}(\vec r)
=
  \begin{pmatrix}
    \sqrt{\beta_L} r_1 & -\sqrt{\beta_N} r_2 \\
    \sqrt{\beta_L} r_2 &  \sqrt{\beta_N} r_1
  \end{pmatrix}
$$
with
\begin{equation}
\label{def:beta-2D}
  \beta_L
=
  2 \tau^{-2} D_1 (2\wp+1)
\qquad\qquad
  \beta_N
=
  2 \tau^{-2}D_1 (3-2\wp)
\end{equation}
This case is special, since $R$ (the matrix of some orthogonal basis
whose first vector is $\vec r$; cf.\ Sect.~\ref{ssec:red-dim}) can be
given as a linear function of $\vec r$, namely we chose above $R=(\vec
r, O_{\pi/2} \vec r)$, where $O_{\pi/2}$ stands for rotation by angle
$\pi/2$ in the positive direction.  In general there is no such linear
representation (however for dimensions $d$ that are a power of 2, the
Cayley-Hamilton algebra of dimension $d$ permits to do a similar
rewriting).  Then
$$
  \tilde{B}(\vec r) \ul{\dd\tilde w}
=
  \begin{pmatrix}
    \sqrt{\beta_L} r_1 \dd\tilde w_1 - \sqrt{\beta_N} r_2 \dd\tilde w_2 \\
    \sqrt{\beta_L} r_2 \dd\tilde w_1 + \sqrt{\beta_N} r_1 \dd\tilde w_2
  \end{pmatrix}
=
  \begin{pmatrix}
    \sqrt{\beta_L} \dd\tilde w_1 & -\sqrt{\beta_N} \dd\tilde w_2 \\
    \sqrt{\beta_N} \dd\tilde w_2 &  \sqrt{\beta_L} \dd\tilde w_1
  \end{pmatrix}
  \begin{pmatrix} r_1 \\ r_2 \end{pmatrix}
$$
Thus, if we define
$$
  \tilde\sigma
=
  \begin{pmatrix}
    \sqrt{\beta_L} \dd\tilde w_1 & -\sqrt{\beta_N} \dd\tilde w_2 \\
    \sqrt{\beta_N} \dd\tilde w_2 &  \sqrt{\beta_L} \dd\tilde w_1
  \end{pmatrix} /\dd t
$$
so that
\begin{equation}
\label{eq:sigmatilde-Btilde}
  \tilde\sigma \vec r \,\dd t
=
  \tilde{B}(\vec r) \ul{\dd\tilde w}
\end{equation}
then \eqref{eq:SDE-red1} can be written as the system
\begin{equation}
\label{eq:psep-2D-red}
  \frac{\dd \vec r}{\dd t}
=
  \vec v
,\qquad
  \frac{\dd \vec v}{\dd t}
=
  -\frac{1}{\tau} \vec v + \tilde\sigma \vec r
\end{equation}
On the other hand we see that, if we identify $\RR^2$ to $\CC$ in the
standard way, then $\tilde\sigma$ corresponds to multiplication by
(and we re-employ $\tilde\sigma$ by abuse of notation)
\begin{equation}
\label{def:sigmatilde-complex}
   \tilde\sigma
=
  \sqrt{\beta_L} \,\frac{\dd\tilde w_1}{\dd t}
 + i \sqrt{\beta_N} \,\frac{\dd\tilde w_2}{\dd t}
\end{equation}
We may pass altogether to the complex notation
$$
  \frac{\dd  r}{\dd t}
=
  v
,\qquad
  \frac{\dd v}{\dd t}
=
  -\frac{1}{\tau} v + \tilde\sigma r
$$
where $r$ and $v$ are the complex equivalents of $\vec r$ and $\vec v$
respectively.

We may proceed further analogously to Sect.~\ref{ssec:Anderson-form},
keeping $E$ from \eqref{def:E,V} and defining $V = \tilde\sigma/\tau$
which is now a complex-valued random potential.  Then $\psi =
e^{t/2\tau} r$ verifies
\begin{equation}
\label{eq:Anderson-complex}
  -\frac{\dd^2 \psi}{\dd t^2} + V \psi
=
  E \psi
\end{equation}
similar to \eqref{eq:Anderson-multi}.

On the other hand one may proceed along the lines of
Sect.~\ref{ssec:Analytic-form}, introducing $\tilde z = v/r$.  Then
$z$ verifies $\dd \tilde z / \dd t = -\tilde z^2 -\tilde z/\tau +
\tilde\sigma$ and $z = \tilde z + 1/2\tau$ verifies
\begin{equation}
\label{eq:z-evol}
  \frac{\dd z}{\dd t}
=
  - z^2 + \frac{1}{4\tau^2} + \tilde\sigma
\end{equation}
Note finally (notwithstanding that $r$ is complex)
$$
  \lambda
=
  \langle \frac{\dd}{\dd t} \log r \rangle
=
  \langle \frac{v}{r} \rangle
=
  -\frac{1}{2\tau} + \langle z \rangle
$$

\subsection{An other solvable case}
\label{ssec:complex-Laplace}

Following ideas of \cite{2dlyapShort}, we will now show that in the 2D
case the Lyapunov exponent may be explicitly computed also when
$\beta_L=0$.  The important particularity of the $\beta_L=0$ (i.e.\
$\wp=-1/2$) situation is that the steady-state distribution of $z$ is
concentrated entirely on the complex half-plane with $\Re z >
1/(2\tau)$, due to the fact that on the line $\Re z = 1/(2\tau)$ the
velocity field always points inside the above-mentioned half-plane.
Indeed, writing $z=x+iy=1/(2\tau)+iy$ we have the velocity field
$$
  -E - z^2
=
  \frac{1}{(2\tau)^2} - (\frac{1}{2\tau}+iy)^2
=
  y^2 - \frac{i}{2\tau}y
$$
and $y^2 > 0$ except for $y=0$.

In this particular situation the Laplace transform $F(p) = \langle
\exp(-pz) \rangle$ is well defined for $p \in \RR_+$.  Using the \Ito
formula
$$
  \dd e^{-pz}
=
  -p e^{-pz} \dd z
  +\frac{p^2}{2} e^{-pz} \langle \dd z, \dd z \rangle
$$
and substituting $\dd z$ according to \eqref{eq:z-evol}, we get for
the steady state
$$
  0
=
  \frac{\dd}{\dd t} F(p)
=
  \left\langle
    \left[ (E+z^2)p - \frac{\beta_N}{2} p^2 \right] e^{-pz}
  \right\rangle
=
  p \left[ \partial_p^2 + E - \frac{\beta_N}{2} p \right] F(p)
$$
Since the above is defined only for $p \geq 0$, we may safely divide
by $p$ and retain simply
$$
  0
=
  \left[ \partial_p^2 + E - \frac{\beta_N}{2} p \right] F(p)
$$
At $p = 0$ we have the obvious boundary condition $F(0) = 1$.  For $p$
going to $+\infty$ we need $\limsup |F(p)| < 1$ since in the steady
state $|\exp(-pz)| < 1$ almost surely for $p \geq 0$ (due to $z$ being
almost surely in the positive half-plane).  These two boundary
conditions are in general sufficient to uniquely determine $F(p)$ in
the steady state.

This differential equation can be readily solved (e.g.\ by introducing
$u= p - 2E/\beta_N$ it reduces to the Airy equation $(\partial_u^2 -
(\beta_N/2)u) f = 0$).  The solution that decays (in fact the only one
that does not grow exponentially) when $p \to +\infty$ and that
verifies $F(0)=1$ is
$$
  F(p)
=
  \frac{Ai(-(\frac{\beta_N}{2})^{-2/3} E + (\frac{\beta_N}{2})^{1/3}p)}
       {Ai(-(\frac{\beta_N}{2})^{-2/3} E)}
$$
where $Ai$ stands for the Airy function of first kind.  Finally we
arrive at
$$
  \langle z \rangle
=
  -(\partial F)(0)
=
  -\left(\frac{\beta_N}{2}\right)^{1/3}
   \frac{Ai'(-(\frac{\beta_N}{2})^{-2/3} E)}
        {Ai (-(\frac{\beta_N}{2})^{-2/3} E)}
$$
Introducing $c = 1/[4\tau^2(\beta_N/2)^{2/3}] =
-(\frac{\beta_N}{2})^{-2/3} E$ (note that preceding definition of $c$
was with $\beta_L$ instead of $\beta_N$!), we have for the Lyapunov
exponent
\begin{equation}
\label{eq:lambda-Ai}
  \lambda
=
  \frac{1}{2\tau}
  \left[
    -1 - c^{-1/2} \frac{Ai'(c)}{Ai(c)}
  \right]
\end{equation}

\subsection{Decay of Lyapunov exponent at large Stokes numbers}

Let us introduce the adimensionalized Stokes time
$$
  \mathit{St}
=
  D_1 \tau
$$
which is nothing else than the Stokes number for the case of a smooth
Kraichnan velocity field.  We are interested here by the decay of the
adimensionalized Lyapunov exponent $\lambda/D_1$ when the Stokes
number $\mathit{St}$ goes to infinity.

Numerical results (see subsection \ref{ssec:naive-sim} and figure
\ref{fig:Lyap-plots}) seem to show us that the Lyapunov exponent
$\lambda$ of pair separation varies monotonically with the
compressibility degree $\wp$ of the advecting flow, notably the larger
$\wp$ (i.e.\ the more compressible the flow) the smaller is $\lambda$.
Thus the two extreme cases of $\wp=-1/2$ and $\wp=3/2$ bound the
other, intermediate cases.

It is then noteworthy that for large $\tau$ the exponents given by
both \eqref{eq:lambda-AiBi} and \eqref{eq:lambda-Ai} decay at the same
rate.  Indeed, we have $\beta_L, \beta_N \sim D_1 \tau^{-2}$ so in
both cases $c \sim \mathit{St}^{-2/3}$, and since $Ai(c)$, $Bi(c)$,
$Ai'(c)$ and $Ai'(c)Ai(c) + Bi'(c)Bi(c)$ all have finite non-zero
limits when $c\to0$, we have the behavior $\lambda/D_1 \sim
\mathit{St}^{-2/3}$ in both cases, so this should be also the general
situation.

An other possible way to arrive at this result is by taking $\tau$ to
$\infty$ while fixing $\tilde\sigma$ in \eqref{eq:z-evol}.  At the
limit, when $1/\tau = 0$, we have simply $z = \tilde z$ and
\eqref{eq:z-evol} reduces to
$$
  \frac{\dd z}{\dd t}
=
  -z^2 + \tilde\sigma
$$
It would then be enough to see that this evolution equation leads to a
(unique) stationary state with $\langle z \rangle$ finite and
non-zero, and that this is indeed the $\tau\to\infty$ limit of
$\langle z \rangle$ (i.e.\ commutation of $t\to\infty$ and
$\tau\to\infty$ limits).

Indeed, since $\tilde\sigma$ is fixed, this means that in fact
$\beta_L, \beta_N$ are fixed, so that (recall \eqref{def:beta-2D})
$D_1 \sim \tau^2$ and $\mathit{St} \sim \tau^3$.  Since we suppose
that for fixed $\sigma$ the Lyapunov exponent $\lambda$ goes to a
constant, the adimensionalized Lyapunov exponent $\lambda/D_1$ behaves
like $D_1^{-1} \sim \tau^{-2} \sim \mathit{St}^{-2/3}$:
$$
  \frac{\lambda}{D_1}
\mathop\sim\limits_{\mathit{St}\to\infty}
  \mathit{St}^{-2/3}
$$

\section{Other solvable 2D flows}
\label{sec:other-sol}

The idea in this section is that the above two special cases of
solvable situations can be slightly generalized.  The cases we could
solve were when $\tilde\sigma$ of \eqref{def:sigmatilde-complex} was
either $\sqrt{\beta_L} \,\dd\tilde w_1/\dd t$ or $i \sqrt{\beta_N}
\,\dd\tilde w_2/\dd t$.  In fact we will now deal with the more
general case of
\begin{equation}
\label{eq:gen-sigmatilde}
  \tilde\sigma
=
  \sqrt{\beta} \,\dd\tilde w/\dd t
\end{equation}
for $\beta \in \CC$ (and $\dd\tilde w$ is the standard white noise).

By analogy with Sect.~\ref{sec:2D}, where $\beta_L + \beta_N =
8D_1/\tau^2$ (from \eqref{def:beta-2D}), we define here the
characteristic inverse time-scale of the flow as
$$
  D_1
=
  \frac{\tau^2}{8} |\beta|
$$

\subsection{Solvable flows}

Let us start by finding the 2D flows which lead to such
$\tilde\sigma$.  Noticing that one has (analogously to
\eqref{eq:BBT-stoch})
$$
  \tilde B_{ik}(\vec r) \tilde B_{jk}(\vec r)
=
  \langle
    \tilde B_{ik} \dd \tilde w_k \tilde B_{jl} \dd \tilde w_l
  \rangle / \dd t
$$
and using \eqref{eq:B-Btilde} to re-express the left hand side and
\eqref{eq:sigmatilde-Btilde} to re-express the right hand side, we get
\begin{equation}
\label{eq:C-sigmatilde}
  \tau^{-2} C_{ik,jl} r_k r_l
=
  \langle
    \tilde\sigma_{ik} r_k \dd t, \tilde\sigma_{jl} r_l \dd t
  \rangle \,/\dd t
\end{equation}
so if we define
$$
  \tilde C_{ik,jl}
=
  \langle
    \tilde\sigma_{ik} \dd t, \tilde\sigma_{jl} \dd t
  \rangle \,/\dd t
$$
then \eqref{eq:C-sigmatilde} can be written
\begin{equation}
\label{eq:C-Ctilde}
  \tau^{-2} C_{ik,jl} r_k r_l
=
  \tilde C_{ik,jl} r_k r_l
\end{equation}

Considering the complex $\tilde\sigma$ of \eqref{eq:gen-sigmatilde} as
a real linear transformation acting on $\CC = \RR^2$, its matrix is
written (also $\tilde\sigma$ by abuse of notation), with help of the
notation $\sqrt{\beta} = \alpha = \alpha_L + i \alpha_N$ ($\alpha \in
\CC$, $\alpha_L, \alpha_N \in \RR$)
$$
  \tilde\sigma
=
  \begin{pmatrix}
    \alpha_L & -\alpha_N \\
    \alpha_N &  \alpha_L \\
  \end{pmatrix}
  \dd\tilde w / \dd t
$$
In coordinate notation this gives
$$
  \tilde\sigma_{ik}
=
  (\alpha_L \delta_{ik} - \alpha_N \epsilon_{ik})
  \,\dd\tilde w / \dd t
$$
Leading to
\begin{align*}
  \tilde C_{ik,jl}
&=
  (\alpha_L \delta_{ik} - \alpha_N \epsilon_{ik})
  (\alpha_L \delta_{jl} - \alpha_N \epsilon_{jl})
\intertext{and using \eqref{eq:epsilon-delta}, this further gives}
&=
  \alpha_L^2 \delta_{ik} \delta_{jl}
 +\alpha_N^2 (\delta_{ij} \delta_{kl} - \delta_{il} \delta_{jk})
 -\alpha_L \alpha_N
    (\epsilon_{ik} \delta_{jl} + \epsilon_{jl} \delta_{ik})
\end{align*}

With the above form for $\tilde C_{ik,jl}$ and using the most general
form respecting isotropy for $C_{ik,jl}$ (cf.\
Sect.~\ref{ssec:broken-sym}), that is:
$$
  C_{ik,jl}
=
  a \delta_{ij} \delta_{kl} +
  b \delta_{ik} \delta_{jl} +
  c \delta_{il} \delta_{jk} +
  d (\epsilon_{ik} \delta_{jl} + \epsilon_{jl} \delta_{ik})
$$
with arbitrary $a,b,c,d \in \RR$, equation \eqref{eq:C-Ctilde} becomes
\begin{multline*}
  \tau^{-2} (a \delta_{ij} r^2 + (b + c) r_i r_j
             + d (\epsilon_{ik} r_j r_k + \epsilon_{jk} r_i r_k))
=\\
  \alpha_N^2 \delta_{ij} r^2 + (\alpha_L^2 - \alpha_N^2) r_i r_j
  -\alpha_L \alpha_N (\epsilon_{ik} r_j r_k + \epsilon_{jk} r_i r_k)
\end{multline*}
Coefficients are identified
$$
  a = \alpha_N^2
\qquad
  b+c = \alpha_L^2 - \alpha_N^2
\qquad
  d = -\alpha_L \alpha_N
$$
and since only the sum of $b$ and $c$ is determined, let us write $b =
\alpha_L^2 - s$, $c = s-\alpha_N^2$ for some $s \in \RR$.  The
positivity conditions \eqref{eq:pos1}, \eqref{eq:pos2} give
\begin{gather*}
  s = a+c \geq 0 \\
  -s = a+b - \sqrt{(b+c)^2 + 4d^2} \geq 0
\end{gather*}
meaning that the only possibility is to take $s=0$ giving
\begin{align*}
  C_{ik,jl}
&=
  \tau^2 \left[
    \alpha_L^2 \delta_{ik} \delta_{jl}
   +\alpha_N^2 (\delta_{ij} \delta_{kl} - \delta_{il} \delta_{jk})
   -\alpha_L \alpha_N
      (\epsilon_{ik} \delta_{jl} + \epsilon_{jl} \delta_{ik})
  \right]
\\
&=
  \tau^2
  (\alpha_L \delta_{ik} - \alpha_N \epsilon_{ik})
  (\alpha_L \delta_{jl} - \alpha_N \epsilon_{jl})
\end{align*}
Such $C_{ik,jl}$ can occur only in a flow (cf.\
Sect.~\ref{ssec:broken-sym}) where parity invariance is broken because
$\epsilon_{ik} \delta_{jl} + \epsilon_{jl} \delta_{ik}$ appears, and
spatial homogeneity is broken because $\delta_{ik} \delta_{jl}$ and
$\delta_{il} \delta_{jk}$ have different coefficients (namely $\tau^2
\alpha_L^2$ and $-\tau^2 \alpha_N^2$, which are equal only in the case
$\tau \alpha_L = \tau \alpha_N = 0$, i.e.\ $C_{ik,jl} = 0$: a very
trivial case).

\subsection{Calculation of the Lyapunov exponent}

This is very much like in Sect.~\ref{ssec:complex-Laplace}.

\begin{figure}[ht]
\center
\includegraphics{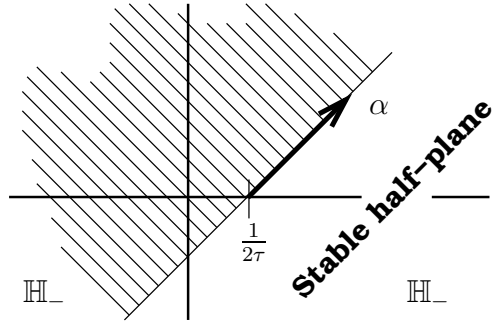}
\caption{The complex plane of $z$ is represented.  There is diffusion
  only in a single direction, parallelly to $\alpha = \sqrt\beta$.
  The dynamics of $z$, governed by \eqref{eq:z-evol} with
  $\tilde\sigma$ as in \eqref{eq:gen-sigmatilde}, is such that $z$ can
  cross the boundary line between the hashed half-plane (unstable) and
  the clear half-plane (stable) only moving out from the hashed
  half-plane and into the clear one, whereby asymptotically the
  distribution of $z$ will be supported entirely on the stable
  half-plane.  $\HH_-$ is the lower half-plane.}
\end{figure}

Let us first establish the existence of a stable half-plane for $z$,
by which we mean one on whose boundary line the drift velocity of $z$
points always towards the interior of the half-plane.  We may suppose,
without loss of generality, that $0 \leq \arg \beta \leq \pi$ and
in consequence $0 \leq \arg \alpha \leq \pi/2$.  We then claim that
the half-plane
$$
  \{ z: \frac{z-\frac{1}{2\tau}}{\alpha} \in \HH_-\}
$$
with $\HH_-$ denoting the complex half-plane with negative imaginary
part is such.  Note that this half-plane may be parametrised as
\begin{equation}
\label{eq:stab-hp-param}
  \frac{1}{2\tau} + \mu \alpha - i \nu \alpha
\qquad
  \mu\in\RR, \nu \in \RR_+
\end{equation}
In particular its boundary line is parametrised as $1/2\tau +
\mu \alpha$ with $\mu\in\RR$, and $-i \alpha$ is perpendicular to the
the boundary line and points to the interior of the stable half-plane.

What we need to show is that, on the boundary line, the scalar product
of the drift of $z$ with $-i\alpha$ (considered as a vector of
$\RR^2$) is positive.  Remark that, viewed as vectors of $\RR^2$, the
scalar product of two complex numbers $x$ and $y$ may be written as
$\Re(x \bar y)$.  Then we have for $z = 1/2\tau + \mu \alpha$
\begin{multline*}
  \Re [(-E-z^2) (\overline{-i\alpha})]
=
  \Re [(-\frac{1}{\tau} \mu\alpha - \mu^2\alpha^2 )
       (\overline{-i\alpha})]
\\
=
  \Re [-\frac{1}{\tau} \mu i |\alpha|^2 - \mu^2|\alpha|^2 i\alpha]
=
  \mu^2|\alpha|^2 \Re(-i\alpha)
\end{multline*}

The idea is again that in the stationary state the distribution of $z$
will be supported by the stable half-plane, being identically zero
outside of it.  In this case the Laplace transform $F(p) = \langle
\exp(pz) \rangle$ is well defined for $p \in (i\alpha)^{-1} \RR_+$
since then $\Re (pz) \leq \Re (1/2\tau p)$ as can be easily read off
from \eqref{eq:stab-hp-param}, making $|\exp(pz)|$ uniformly bounded
in $z$ for fixed $p$.

As previously, in the steady state we may write
$$
  0
=
  \frac{\dd}{\dd t} F(p)
=
  \left\langle
    \left[ -(E+z^2) p + \frac{\beta}{2} p^2 \right] e^{pz}
  \right\rangle
=
  p (-\partial_p^2 - E + \frac{\beta}{2} p) F(p)
$$
We may divide the above by $p$ without loss of information, since at
$p=0$ we have the boundary condition $F(0) = 1$, and as $|p| \to
\infty$ with $p \in (i\alpha)^{-1} \RR_+$, the $F(p)$ has to decay,
and these two conditions determine uniquely the solution.
\begin{equation}
\label{eq:F-stat-ed}
  (\partial_p^2 + E - \frac{\beta}{2} p) F(p)
=
  0
\end{equation}
This equation is related to the Airy equation, if we introduce the new
variable $u = p-2E/\beta$:
$$
  (\partial_u^2 - \frac{\beta}{2} u) f(u)
=
  0
$$
of which two independent solutions are (formula 10.4.1 in
\cite{MR1225604}) $Ai((\beta/2)^{1/3} u)$ and $Ai(e^{2\pi
i/3}(\beta/2)^{1/3} u)$, where $Ai$ denotes the Airy function of first
kind, and $(\beta/2)^{1/3}$ is taken with the standard definition of
the root, (i.e.\ real positive on the positive real axis and having
branch cut along the negative real axis).

For large argument, to lowest order, the asymptotic expansion of the Airy
function $Ai$ reads (formulae 10.4.59 of \cite{MR1225604})
\begin{equation}
\label{eq:Ai-asymp}
  Ai(\zeta)
=
  \frac{1}{2 \pi^{1/2} \zeta^{1/4}} e^{-\frac{2}{3} \zeta^{3/2}}
  (1 + O(\zeta^{-3/2}))
\qquad (|\arg \zeta| < \pi)
\end{equation}

Let us apply the above with $\zeta = (\beta/2)^{1/3} u =
(\beta/2)^{1/3} (p-2E/\beta)$.  Since $p \in (i\alpha)^{-1} \RR_+$ and
$\alpha = \sqrt{\beta}$ (with standard definition of the root), when
$|p| \to \infty$ we have $\arg u \to \arg p = -(\arg\beta)/2-\pi/2$,
whence $\arg\zeta \to -(\arg\beta)/6-\pi/2$, so that for large enough
$|p|$ we have $-\pi/2 \geq \arg\zeta \geq -2\pi/3$ (in particular
$|\arg \zeta| < \pi$), and $-3\pi/4 \geq \arg(\zeta^{3/2}) \geq -\pi$.
This implies, when $|p| \to \infty$, that $\Re (\zeta^{3/2}) < 0$, and
since $|\zeta| \sim (|\beta|/2)^{1/3} |p|$, we get from
\eqref{eq:Ai-asymp} that $|Ai((\beta/2)^{1/3} (p-2E/\beta))| \to
\infty$.

Repeating the same treatment with $\zeta = e^{2\pi i/3}
(\beta/2)^{1/3} u = e^{2\pi i/3} (\beta/2)^{1/3} (p-2E/\beta)$, when
$|p| \to \infty$ we have $\arg\zeta \to -(\arg\beta)/6+\pi/6$, so that
for large enough $|p|$ we have $0 \leq \arg\zeta \leq \pi/6$ (in
particular $|\arg \zeta| < \pi$), and $0 \leq \arg(\zeta^{3/2}) \leq
\pi/4$.  This implies, when $|p| \to \infty$, that $\Re (\zeta^{3/2})
> 0$, and since $|\zeta| \sim (|\beta|/2)^{1/3} |p|$, we get from
\eqref{eq:Ai-asymp} that $|Ai(e^{2\pi i/3} (\beta/2)^{1/3}
(p-2E/\beta))| \to 0$.

Hence the unique solution of \eqref{eq:F-stat-ed} verifying the
appropriate boundary conditions at 0 and at infinity is
$$
  F(p)
=
  \frac{Ai(e^{2\pi i/3} (\beta/2)^{1/3} (p-2E/\beta))}
       {Ai(e^{2\pi i/3} (\beta/2)^{1/3} (-2E/\beta))}
$$

Remark that $\langle z \rangle = \partial_p F(p)|_{p=0}$, giving
\begin{equation}
\label{eq:z-mean-1}
  \langle z \rangle
=
  e^{2\pi i/3} (\beta/2)^{1/3}
  \frac{Ai'(e^{2\pi i/3} (\beta/2)^{1/3} (-2E/\beta))}
       {Ai (e^{2\pi i/3} (\beta/2)^{1/3} (-2E/\beta))}
\end{equation}
We see (if we evaluate the above formula for non-real $\beta$), that
$\langle z \rangle$ is in general complex valued.  By analogy with
Sect.~\ref{sec:2D} we define
$$
  \lambda
=
  \langle z \rangle - \frac{1}{2\tau}
$$
The real part of $\lambda$ once again gives the exponential growth
rate of the pair separation $\vec r$, while the imaginary part of
$\lambda$ gives the average rotation speed of $\vec r$.  In particular
this mean angular velocity is in general non-zero.

Eq.~\eqref{eq:z-mean-1} may be further simplified by noticing that
$e^{2\pi i/3} (\beta/2)^{1/3} (-2E/\beta) = -E e^{2\pi i/3}
(\beta/2)^{-2/3}$, and $|e^{2\pi i/3} (\beta/2)^{-2/3}| =
|\beta/2|^{-2/3}$ and $\arg(e^{2\pi i/3} (\beta/2)^{-2/3}) = 2\pi/3 -
(2/3)\arg\beta = (2/3) (\pi - \arg\beta)$, and since $0 \leq \arg\beta
\leq \pi$ we have $\pi - \arg\beta = \arg(-1/\beta) =
\arg((-\beta)^{-1})$ so that $\arg(e^{2\pi i/3} (\beta/2)^{-2/3}) =
\arg((-\beta)^{-2/3})$.  From this we conclude that $e^{2\pi i/3}
(\beta/2)^{-2/3} = (-\beta/2)^{-2/3}$.  Along the same lines,
$\arg(e^{2\pi i/3} (\beta/2)^{1/3})= 2\pi/3 + (\arg\beta)/3 =
(\arg\beta - \pi)/3 + \pi$.  However we have to pay attention here
since $\arg\beta-\pi = \arg(-\beta)$ is true only for $0 < \arg\beta
\leq \pi$ but {\em not} for $\arg\beta = 0$, so to sum it up
$\arg\beta-\pi = \arg(-\beta)$ is true only for $\beta\notin\RR_+$.
Thus, {\em except for the case of $\beta$ real positive} we have
$(\arg\beta - \pi)/3 + \pi = \arg(-(-\beta)^{1/3})$ so that $e^{2\pi
i/3} (\beta/2)^{1/3} = -(-\beta/2)^{1/3}$.  Thus we also have
\begin{equation}
\label{eq:z-mean-2}
  \langle z \rangle
=
  -(-\beta/2)^{1/3}
  \frac{Ai'(-E (-\beta/2)^{-2/3})}
       {Ai (-E (-\beta/2)^{-2/3}))}
\qquad\qquad
  \beta \notin \RR_+
\end{equation}
Note that throughout we have supposed $\beta$ to be in the upper
complex half-plane.  However this last formula has the interesting
feature that it is real, so that substituting $\beta$ with its complex
conjugate $\bar\beta$, we also get for $\langle z \rangle$ the
conjugate value, and from simple symmetry considerations with respect
to the real axis, this is the correct answer.  Thus formula
\eqref{eq:z-mean-2} applies for {\em all} $\beta \in \CC \setminus
\RR_+$.

To understand the problem with $\beta \in \RR_+$, note that in this
particular case, when the diffusion is parallel to the real axis,
there are {\em two} stable half-planes: the lower and the upper
complex half-planes.  The equilibrium distribution of $z$ depends on
the initial distribution of $z$, since $z$ cannot cross the real
line.  As we have already showed, for $z$ starting from the lower
half-plane (this was only implicit in our considerations, through the
fact that we considered the stable half-plane to be the lower one) we
arrive at formula \eqref{eq:z-mean-1}.  Starting from the upper
half-plane we arrive at its complex conjugate.  Hence the natural
branch cut in \eqref{eq:z-mean-2} for $\beta \in \RR_+$.

We recover the two solvable cases presented in
Sect.~\ref{ssec:sol-case-1} and Sect.~\ref{ssec:complex-Laplace}.  The
former one, with $\beta_N = 0$ (hence diffusion only in the real
direction), corresponds to $\beta = \beta_L \in \RR_+$.  Using
\eqref{eq:z-mean-1} and the identity (formula 10.4.9 of
\cite{MR1225604})
$$
  Ai(e^{2\pi i/3} \zeta)
=
  \frac{1}{2} e^{\pi i/3} (Ai(\zeta) - i Bi(\zeta))
$$
leading also to the identity
$$
  Ai'(e^{2\pi i/3} \zeta)
=
  e^{-2\pi i/3} \frac{\dd}{\dd\zeta} Ai(e^{2\pi i/3} \zeta)
=
  \frac{1}{2} e^{-\pi i/3} (Ai'(\zeta) - i Bi'(\zeta))
$$
we have
$$
  \langle z \rangle
=
  (\beta_L/2)^{1/3}
  \frac{Ai'(-E  (\beta_L/2)^{-2/3}) - i Bi' (-E (\beta_L/2)^{-2/3})}
       {Ai (-E (\beta_L/2)^{-2/3}) - i Bi (-E (\beta_L/2)^{-2/3})}
$$
Introduce $c = -E (\beta_L/2)^{-2/3} = 1/[4\tau^2(\beta_L/2)^{2/3}]$,
and multiply numerator and denominator in the above displayed formula
by the complex conjugate of the denominator, finally use the Wronskian
identity (formula 10.4.10 of \cite{MR1225604}, see Index of Notations
p.~1045 of the cited source for sign convention for the Wronskian)
$$
  Ai(\zeta) Bi'(\zeta) - Ai'(\zeta) Bi(\zeta)
=
  \pi^{-1}
$$
in order to get
$$
  \langle z \rangle
=
  \frac{1}{2\tau} c^{-1/2}
  \frac{Ai'(c)Ai(c) + Bi'(c)Bi(c) - i \pi^{-1}}
       {Ai^2(c) + Bi^2(c)}
$$

So what is the meaning of this imaginary part ?  Recall that our
initial assumption was that $z$ is inside the stable half-plane at
asymptotically long times, and as a limiting case of $\arg\alpha
\searrow 0$ this half-plane is that of negative imaginary parts.  The
asymptotic distribution of $z$ is indeed supported in this half-plane
if the initial $z$ is in it.  If however we start with $z$ initially
distributed with probability $1/2$ on each side of the real axis, or
if we add infinitesimally small diffusion in the imaginary direction,
then we get in the upper and lower half-planes distributions which are
mirror images of one-another, and in the lower half-plane it is $1/2$
times that of the case when there was nothing in the upper half-plane.

That $\langle z \rangle$ does have the imaginary part predicted above
can be readily verified in simulations (see
Fig.~\ref{fig:Lyap-Li-plots}, the $\beta = 1$ curve), even if we use
the scheme of infinitesimal diffusion in the direction of the
imaginary axis, by evaluating $\langle |\Im z| \rangle$ and confirming
the relation
$$
  \langle |\Im z| \rangle
=
  \frac{1}{2\tau} c^{-1/2}
  \frac{\pi^{-1}}
       {Ai^2(c) + Bi^2(c)}
 $$

We also recover the case $\beta_L = 0$ (hence diffusion only in the
imaginary direction) corresponding to $\beta = -\beta_N \in \RR_-$.
Here we use \eqref{eq:z-mean-2} and get
$$
  \langle z \rangle
=
  -(\beta_N/2)^{1/3}
  \frac{Ai'(-E (\beta_N/2)^{-2/3})}
       {Ai (-E (\beta_N/2)^{-2/3}))}
$$
or, introducing $c = -E (\beta_N/2)^{-2/3} = 1/[4\tau^2
(\beta_N/2)^{2/3}]$
$$
  \langle z \rangle
=
  -\frac{1}{2\tau} c^{-1/2}
  \frac{Ai'(c)}
       {Ai (c)}
$$

\section{Numerical results}

In this section we focus mainly on the $d=2$ dimensional case.

\subsection{Naive simulation of inertial particle separation}
\label{ssec:naive-sim}

In order to get the Lyapunov exponent of pair separation, the most
straightforward thing to do is to simulate the reduced equation
\eqref{eq:psep-2D-red}.  Note that this is better than simulating
\eqref{eq:z-evol}, since at moments when particles nearly collide
(happens when particles are moving towards each other with small impact
parameter (compared to $\tau v$)), the quotient $v/r = z-1/2\tau$ turns
very fast, which would require additional handling in the code,
whereas in fact $v$ evolves little at near collision since $v \gg
r/\tau$ and $r$ evolves linearly for which no special care is needed
in the simulations.

The convergence of the Lyapunov exponent is quite slow, one can hope
for a 0.1\% relative precision at most, which is not enough to try to
guess a functional form through Plouff's inverter, but more than
enough to disprove some claims of formulae.

Note also that for $\tau$ small the system is stiff and convergence
gets worse.  The equation of evolution should be solved analytically
between two time-steps, with taking for $\vec r$ on the right-hand
side its value at the beginning of the time interval.  Indeed in this
case we have a linear (multi-dimensional) equation with
time-independent homogeneous term (only the inhomogeneous term contains
the driving noise) which can be explicitly solved, and we get an
effective driving term whose statistics can be computed.  This is
treated in subsection \ref{ssec:stiffness}.

Numerical simulations were done in series with fixed $\beta_L$,
$\beta_N$ (with additionally fixing $\beta_L + \beta_N = 1$; note the
identity $\beta_L + \beta_N = 8D_1/\tau^2$ leading to $D_1 = \tau^2/8$
with our choice) instead of fixed $D_1$ (note that $\wp$ is fixed by
the quotient of $\beta_L$, $\beta_N$).  However we are only interested
in the the adimensionalized Lyapunov exponent $\lambda/D_1$ in
function of the Stokes number $\mathit{St} = D_1 \tau$.  Simulations
were done for values of $\tau$ between $0.1$ and $4$, which, according
to the above, correspond to Stokes numbers between
$1.25\!\cdot\!10^{-4}$ and $8$.  Also, for reasons of better numerical
stability, instead of taking $\beta_N = 0$, we used in that case
$\beta_N = 10^{-6}$.  Total time of simulation was $5\!\cdot\!10^5$
with time-step $10^{-4}$.

The plot of numerical results, at the end of the paper, has been cut
up into three pieces ($[0,0.1]$, $[0.1,1]$ and $[1,7.5]$ that are
shown on different scales).  The vertical axis is $\lambda/D_1$ the
adimensionalized Lyapunov exponent.  The curves are ordered in the
same top to bottom order as the keys.  Solid line curves are the
analytical solutions, dashed lines are just linear interpolation of
numerical data.  Data itself is represented by boxes, either open or
closed.

\subsection{Estimates of \texorpdfstring{$\vec v$}{v} and of the
  change of \texorpdfstring{$\vec r$}{r} during a time interval
  \texorpdfstring{$\Delta t$}{\textDelta t}}
\label{ssec:estimates}

Let us estimate first $\vec v$ and then the jump in $\vec r$ during
some time interval $\Delta t$.  We suppose $|\vec r| \approx 1$ at the
beginning of the time interval, without loss of generality, since the
system under study is scale invariant.  The case we are interested in
is when $\Delta t$ is the time-step of our simulation, but here we
talk about the ``real'' $\vec v$ and $\vec r$, not the simulated ones.
Hence we only consider the case where $\Delta t \ll \ D_1^{-1}$.

The forcing of $\vec v$ is a white noise with intensity proportional to
$\tau^{-1} \sqrt{D_1}$ and the ``memory'' of $\vec v$ is of order
$\tau$, so that $|\vec v| \sim \tau^{-1} \sqrt{D_1} \sqrt{\tau} =
\sqrt{D_1/\tau}$.

As for $\vec r$, we need to distinguish the cases when $\tau < \Delta
t$ and when $\tau > \Delta t$.  The time derivative of $\vec r$ is
$\vec v \sim \sqrt{D_1/\tau}$.  Now if $\tau < \Delta t$ then we can
say that $\vec v$ is correlated over a time $\tau$ so that during each
interval of time of length $\tau$, the vector $\vec r$ changes by
approximately $\tau \sqrt{D_1/\tau} = \sqrt{D_1 \tau}$.  There are
$\Delta t / \tau$ such independent changes which gives a total change
of $\sqrt{D_1 \tau} \sqrt{\Delta t / \tau} = \sqrt{D_1 \Delta t}$,
just as if $\vec r$ underwent a stochastic diffusion with diffusion
constant not depending on $\tau$.

In the other case of $\tau > \Delta t$ the velocity is correlated over
times larger than $\Delta t$ so that the change of $\vec r$ is
approximately $|\vec v| \Delta t \sim \sqrt{D_1/\tau} \Delta t$.  For
later use, we remark that this latter can be estimated, due to the
hypothesis $\tau > \Delta t$, as $\sqrt{D_1/\tau} \Delta t \leq
\sqrt{D_1/\Delta t} \Delta t = \sqrt{D_1 \Delta t}$, so that in this
case also $\vec r$ changes by at most $\sqrt{D_1 \Delta t}$.

\subsection{Integrating \texorpdfstring{$\dd \log r^2$}{d log
    r\texttwosuperior}}

To obtain numerically the Lyapunov exponent $\lambda$, we essentially
make use of the formula
$$
  \lambda
=
  \left\langle \frac{\dd\log r^2}{2 \, \dd t} \right\rangle
$$

However, as stated above, the change $\Delta \vec r$ of $\vec r$
during some time interval $\Delta t$ can be bounded, in general, only
as $\Delta \vec r < \sqrt{D_1 \Delta t}$ (but at least this bound is
uniform in $\tau$).  This is fine, but it means that to calculate the
change of a non-linear function $f$ of $\vec r$ we have to develop $f$
to second order in $\vec r$, which means in other words that we need
to use the \Ito formula.

In practice, we have
\begin{equation}
\label{eq:log-Ito}
  \dd \log r^2
=
    \frac{1}{r^2} 2 r_i \,\dd r_i
  + \Big( -\frac{2}{r^4} r_i r_j + \frac{1}{r^2} \delta_{ij} \Big)
      \,\dd r_i \,\dd r_j 
\end{equation}
and this is the formula to use in simulations to get a correct result
(in the simulations it appeared clearly that for $\tau < \Delta t$ the
first-order formula gives completely false results).

\subsection{Overcoming stiffness}
\label{ssec:stiffness}

For small values of the Stokes time $\tau$ the system has a large
parameter (v.g.\ $\tau^{-1}$), so that it is stiff.  In principle one
needs to take a time-step $\Delta t$ to integrate it that is smaller
by one or several orders of magnitude than $\tau$.  However in the
$\tau \to 0$ limit the resulting process for $\vec r$ is still well
defined, as a diffusion process, which may still be integrated with a
finite time-step $\Delta t$ which only has to be small compared to the
inverse diffusion coefficient $D_1^{-1}$.  Can we devise an integration
scheme that doesn't need $\Delta t \to 0$ as $\tau \to 0$ and which is
correct both for $\tau < \Delta t$ and for $\tau > \Delta t$, given
$\Delta t \ll D_1^{-1}$ ?

The answer is affirmative as we now expose.  Since in the $\tau \to 0$
limit the ``fast'' variable is $\vec v$ we will integrate its
evolution over one time-step say between 0 and $t=\Delta t$ (for ease
of notation we will use here $t$ instead of the more clumsy $\Delta
t$), analytically.  This is possible if we keep in the evolution
equation $\vec r$ fixed to its value $\vec r(0)$ at the beginning of
the time-step, since then the system becomes linear with a constant
evolution matrix and some time-dependent forcing:
\begin{equation}
\label{eq:exp-const}
  \dd \begin{pmatrix} \vec r \\ \vec v \end{pmatrix}
=
    \begin{pmatrix} 0 & 1 \\ 0 & -\frac{1}{\tau} \end{pmatrix}
      \begin{pmatrix} \vec r \\ \vec v \end{pmatrix} \dd t
  + \begin{pmatrix} 0 \\ \dd S \vec r(0) \end{pmatrix}
\end{equation}
See Appendix \ref{app:const-approx} for justification of this
approximation.

Let us start out with the study of a generic system of this form
\begin{equation}
\label{eq:gen-exp-const}
  \dd \vec X
=
  A \vec X \dd t + \dd \vec B_t
\end{equation}
Its solution is
\begin{equation}
\label{eq:int-X}
  \vec X(t)
=
    e^{tA} \vec X(0)
  + \int_0^t e^{(t-s)A} \dd \vec B_s
\end{equation}
We thus have a deterministic part and a stochastic one which is linear
in the increments of the Brownian driving process $B_s$, in other
words the stochastic part is in the first chaos of $B_s$, hence it is
a mean zero Gaussian random variable.  Also since on (sub-intervals
of) different time-step intervals the increments of the driving $B_s$
are independent, we really need to know only the covariance of the
stochastic part on any given time-step:
$$
  \left\langle
    \left( \int_0^t e^{(t-s)A} \dd \vec B_s \right) \otimes
    \left( \int_0^t e^{(t-s)A} \dd \vec B_s \right)
  \right\rangle
$$
Let us call $K(s) = \exp(sA)$, and notice that for a vector $\vec
u$ we have the equivalent writings $\vec u \otimes \vec u = \vec u
\vec u^T$.  Introduce the equal-time correlation matrix $\chi$ such
that $\langle \dd \vec B_s \otimes \dd \vec B_{s'} \rangle = \chi
\delta(s-s') \, \dd s \dd s'$.  Then the above displayed formula
simplifies to
\begin{align}
  \int_0^t \int_0^t
    K(t-s) \langle \dd \vec B_s \otimes \dd \vec B_{s'} \rangle K^T(t-s')
&=
\\
\label{eq:cov-DX}
  \int_0^t \int_0^t K(t-s) \chi K^T(t-s') \delta(s-s') \, \dd s \dd s'
&=
  \int_0^t  K(t-s) \chi K^T(t-s) \, \dd s
\end{align}
At this point one needs to write the above correlation matrix as
$UU^T$ for some matrix $U$.  This is most easily done by way of the
Cholesky decomposition which is simply the prescription of $U$ being
triangular (lower or upper), because the coefficients of $U$ can then
be recursively computed by using only algebraic operations and square
roots.  This is algorithmically much simpler than if for example we
tried to diagonalize the correlation matrix to take its symmetric
square-root.

In our case it is easily computed that
$$
  K(t-s)
=
  \begin{pmatrix} 1 & \tau (1-e^{-\frac{t-s}{\tau}}) \\
                  0 & e^{-\frac{t-s}{\tau}}
  \end{pmatrix}
$$

but note that this a block-matrix notation where each block is
proportional to the $2 \times 2$ identity matrix.  As for $\chi$, we
can see it also as a $2 \times 2$ block-matrix of $2 \times 2$
matrices, whose only non-zero block is the lower-right one, and in fact
$$
  \chi
=
  \begin{pmatrix} 0 & 0 \\
                  0 & \tilde B(\vec r(0)) \tilde B^T(\vec r(0))
  \end{pmatrix}
$$

Since every block of $K(s)$ (each being a scalar matrix) commutes with
the non-zero block of $\chi$, we can start by evaluating
\begin{align*}
  L(t,\tau)
&=
  \int_0^t
    K(t-s) \begin{pmatrix} 0 & 0 \\ 0 & 1 \end{pmatrix} K^T(t-s)
  \, \dd s
\\
&=
  \begin{pmatrix}
    \ &
    \tau^2 [ t
            - \frac{\tau}{2} (1-e^{-\frac{t}{\tau}})
                             (3-e^{-\frac{t}{\tau}})
           ] & \quad &
    \frac{\tau^2}{2} (1-e^{-\frac{t}{\tau}})^2
    & \ 
  \\[2ex]
    \ &
    \frac{\tau^2}{2} (1-e^{-\frac{t}{\tau}})^2 & \quad &
    \frac{\tau}{2} (1-e^{-\frac{t}{\tau}}) (1+e^{-\frac{t}{\tau}})
    & \ 
  \end{pmatrix}
\end{align*}
and its Cholesky decomposition can be given as
$$
  U
=
  \begin{pmatrix}
    \ & \tau \sqrt{t-2\tau\tanh\frac{t}{2\tau}} & \quad &
    \tau(1-e^{-\frac{t}{\tau}})\sqrt{\frac{\tau}{2}\tanh\frac{t}{2\tau}}
    & \ 
  \\[2ex]
    \ & 0 & \quad & \sqrt{\frac{\tau}{2}(1-e^{-2\frac{t}{\tau}})} & \ 
  \end{pmatrix}
$$

Finally with the full Cholesky decomposition the driving noise process
we need can be generated as
$$
  (U \otimes \tilde B) \, \ul{\dd\eta}
=
  \begin{pmatrix}
    U_{11} \tilde B & U_{12} \tilde B \\
    0               & U_{22} \tilde B
  \end{pmatrix}
  \begin{pmatrix}
    \dd\eta_1 \\ \dd\eta_2 \\ \dd\eta_3 \\ \dd\eta_4
  \end{pmatrix}
$$
with the $\eta$ independent Gaussian white noises, $\langle \dd\eta_i
\dd\eta_j \rangle = \delta_{ij} \dd t$.  It is interesting to notice
that this is tantamount to using to {\em independent} realizations of
the Kraichnan field ($\tilde B (\dd\eta_1,\dd\eta_2)$ and $\tilde B
(\dd\eta_3,\dd\eta_4)$) to drive the inertial particle process.  It is
not difficult to see that this is the general case (i.e.\ more than
two particles in a non-linear velocity field may be adequately driven
in a finite time-step scheme by two independent realizations of the
fluid velocity field, just as here).

\subsection{Extrapolation from moments of positive even order}

Another possibility is to compute the exponential growth rates of
positive even order moments of the particle separation, which can be
obtained analytically (though implicitly only, as the largest real
root of some polynomial), as described in
Appendix~\ref{app:pos-even-mom}.  Then, the curve which we know only
for even natural numbers has to be extrapolated in order to guess its
derivative at 0.  This is a tough business, and it is hard to get
anything better than 10\% accuracy.

There are several ways of calculating the top eigenvalue.  For small
order $2n$ of the moments (until $n \approx 10$), it is possible to
calculate directly the characteristic polynomial of the matrix and
solve numerically for the largest real root.

For larger $n$ (up to $n \approx 100$) it is more convenient to iterate
the system $x \mapsto Mx$ where $M$ is the matrix of the system.
Since $M$ is a sparse matrix, this is fast and efficient and doesn't
require a lot of memory.

For $n$ even larger it is better to use the more memory intensive
method of iterating $A \mapsto A^2$ with $A_0 = M$, since this way the
power of $M$ computed is in $2^N$ for $N$ iterations, whereas for the
previous method it was just $N$.

\section{On some conjectured formulae in 2D}

In \cite{MR1897722} appeared the following conjectured formula for the
top Lyapunov exponent of inertial particles in a 2D Kraichnan flow as
described above:
\begin{equation}
\label{eq:conj-Piterbarg}
  \frac{\lambda}{D_1}
=
  \Re\left(
    -\frac{1}{2\tau D_1}
    \left[ 1 + c^{-1/2} \frac{Ai'(c)}{Ai(c)} \right]
  \right)
\end{equation}
with
\begin{equation}
\label{eq:conj-c}
  c
=
  [4\tau^2(\beta_N-\beta_L)^{2/3}]^{-1}
=
  [16(1-2\wp)\mathit{St}]^{-2/3}
\end{equation}
This would mean that $\lambda/D_1$ changes with $\wp$ only through
simple rescaling.  Indeed, introducing the function
$$
  F(x)
=
  \Re\left(
    -\frac{8}{x}
    \left[ 1 + x^{\frac{1}{3}}
               \frac{Ai'(x^{-\frac{2}{3}})}
                    {Ai (x^{-\frac{2}{3}})}
    \right]
  \right)
$$
formula \eqref{eq:conj-Piterbarg} is equivalent to
\begin{equation}
\label{eq:conj-scaling}
  \frac{\lambda}{D_1}
=
  (1-2\wp) F(16(1-2\wp)\mathit{St})
\end{equation}
with the only subtlety that $F$ behaves differently for positive and
negative arguments.  In fact, we have for $x>0$
$$
  F(x)
=
    -\frac{8}{x}
    \left[ 1 + x^{\frac{1}{3}}
               \frac{Ai'(x^{-\frac{2}{3}})}
                    {Ai (x^{-\frac{2}{3}})}
    \right]
$$
but
$$
  F(-x)
=
    \frac{8}{x}
    \left[ 1 - x^{\frac{1}{3}}
               \frac{ Ai'(x^{-\frac{2}{3}}) Ai(x^{-\frac{2}{3}})
                     +Bi'(x^{-\frac{2}{3}}) Bi(x^{-\frac{2}{3}})}
                    { Ai^2(x^{-\frac{2}{3}})
                     +Bi^2(x^{-\frac{2}{3}})}
    \right]
$$  

To see that \eqref{eq:conj-Piterbarg} cannot agree with numerical
results, the easiest is to see that \eqref{eq:conj-scaling} would
imply $\lambda = 0$ for $\wp=1/2$ for all values of $\mathit{St}$,
since $F(0)=2$, and clearly this is not the case.

We compare the conjectured formula with effective values.  Notice that
for $\wp=0$ the match is not too bad, but for larger values of $\wp$
it becomes worse.

Later, in \cite{PRL2506021}, it was recognized that
\eqref{eq:conj-Piterbarg} cannot be always correct, but it was
hypothetized that perhaps it could be exact for the $\wp=1/6$ case.
To help compare their results with ours, we give the expressions of
their parameters $\Gamma$, $\epsilon^2$, $\gamma$ and $z$ in function
of our variables:
$$
  \Gamma
=
  \frac{3-2\wp}{1+2\wp}
\qquad\qquad
  \epsilon^2
=
  (1+2\wp) \mathit{St}
\qquad\qquad
  \gamma = \frac{1}{\tau}
\qquad\qquad
  z = c
$$
where $c$ is that of formula \eqref{eq:conj-c}.

They also give a modified formula, which in our notations is
$$
  \frac{\lambda}{D_1}
=
  \chi(\wp) e^{-\frac{1}{6(1+2\wp)\mathit{St}}}
 +(1-2\wp) F(16(1-2\wp)\mathit{St})
$$
where $\chi(\wp)$ is an unknown prefactor, that according to the
authors of \cite{PRL2506021} does not depend on $\mathit{St}$.  But
they only claim that the additional term should be some sort of
leading order correction.

\section{Conclusions}

We succeeded, for certain time decorrelated flows, to find an
analytical expression for the Lyapunov exponent of inertial particle
pair separation.  Other cases had to be dealt with numerically.  It is
however interesting that the analytically tractable cases bear lot of
resemblance with the general case and thus have a predictive power.
For example the $D_1^{-1}\lambda \sim \mathit{St}^{-2/3}$ behavior
was first noticed for the analytically available solution.

The finding of slightly generalized cases of the Anderson localization
problem which permit exact solution could be also of broader interest.

We have described so far only the Lyapunov exponent, but a more general
object, the distribution (in fact large deviations) of finite-time
Lyapunov exponents, contains further useful details of particle pair
separation.  Our work can be extended, following ideas of
\cite{MR1953940}, to gain more information on this object.

Hence, the analogy between inertial particles in a random flow and the
Anderson localization problem (which itself is related to the Edwards
polymer model also known as the weakly self-avoiding Brownian motion
(or random walk)) is fruitful and holds still more promises.

\appendix

\section{Constant \texorpdfstring{$\vec r$}{r} approximation}
\label{app:const-approx}

Instead of using the approximated equation \eqref{eq:exp-const}, we
can start out from the true system
$$
  \dd \begin{pmatrix} \vec r \\ \vec v \end{pmatrix}
=
    \begin{pmatrix} 0 & 1 \\ 0 & -\frac{1}{\tau} \end{pmatrix}
      \begin{pmatrix} \vec r \\ \vec v \end{pmatrix} \dd t
  + \begin{pmatrix} 0 \\ \dd S \vec r(t) \end{pmatrix}
$$
which has the generic form (analogously to \eqref{eq:gen-exp-const})
$$
  \dd \vec X
=
  A \vec X \dd t + \dd \vec B_t(\vec X)
$$
where it is in fact the covariance of $\dd \vec B_t$ that depends on
$\vec X(t)$ as $\langle \dd \vec B_s \otimes \dd \vec B_{s'} \rangle =
\chi(\vec X(s)) \delta(s-s') \, \dd s \dd s'$

What is of interest to us in the numerical simulation is the mean and
covariance of
$$
  \Delta \vec X
=
  \int_0^t \dd \vec X
=
  \vec X(t) - \vec X(0)
$$

We can again write a formula of type \eqref{eq:int-X}
$$
  \vec X(t)
=
    e^{tA} \vec X(0)
  + \int_0^t e^{(t-s)A} \dd \vec B_s(\vec X(s))
$$
but now this is an implicit formula since it contains $\vec X(s)$ for
$s>0$ on the right hand side also.

For the mean displacement we still have immediately
$$
  \langle \Delta \vec X \rangle
=
  (e^{tA}-1) \vec X(0)
$$

For the covariance however we only have, instead of \eqref{eq:cov-DX},
the semi-explicit
$$
  \langle
    [\Delta \vec X  - \langle \Delta \vec X \rangle]
    \otimes
    [\Delta \vec X  - \langle \Delta \vec X \rangle]
  \rangle
=
  \int_0^t  K(t-s) \, \chi(\vec X(s)) \, K^T(t-s) \, \dd s
$$

It is at this point that we use the estimates of subsection
\ref{ssec:estimates}, notably that $\vec X(s) - \vec X(0) \lesssim
\sqrt{D_1 s} \leq \sqrt{D_1 t}$ if we take $|\vec X(0)| \approx 1$.
Then we have a {\em relative} error by replacing $\chi(\vec X(s))$
with $\chi(\vec X(0))$ of order bounded by $\sqrt{D_1 t}$.  This
basically means that in our approximate method we simulate a system
whose diffusion coefficient would fluctuate around the real one, with
a {\em relative} precision of order $\sqrt{D_1 t}$.  Since in the
simulations that are reported here we took $D_1 = \tau^2/8$ and a
time-step of $t=10^{-4}$, we have $\sqrt{D_1 t} \tau \simeq
3.5\,10^{-3} \tau$, and since we used only $\tau < 4$, we have
$\sqrt{D_1 t} \tau \leq 1.5 \!\cdot\! 10^{-2}$.

\section{The \texorpdfstring{$\tau\to0$}{\texttau{} goes to 0} limit}
\label{app:tau-to-0}

When the Stokes time $\tau$ goes to 0, the movement of an inertial
particle solving \eqref{eq:inertial-particles} tends to that of a
passive tracer that follows the velocity field, solving the SDE
\begin{equation}
\label{eq:passive-tracer}
  \dd \vec R(t)
=
  \dd \vec U_t(\vec R(t))
\end{equation}
where we preferred writing $\dd \vec U_t$ instead of $\vec U$ to
clearly mark that the velocity field is a white noise in time.  The
solution of this equation depends, in general, on the convention we
adopt (\Ito or Stratonovich or other), except if the flow is
incompressible or its statistics is locally isotropic.

However, as pointed out in Sect.~\ref{ssec:conv-indiff}, for finite
$\tau$ the evolution of the inertial particle does not depend on the
adopted convention: we don't have any degree of freedom.  Thus, in the
$\tau\to0$ limit we must find also a unique prescription how to
integrate \eqref{eq:passive-tracer}.

The same reasoning can be carried out for individual particle
trajectories and particle pair separation, and since we have mostly
developed formulae for the latter, we will use that setting.  From the
development of $\vec r$ for a small time-step $\Delta t$, as obtained
in subsection \ref{ssec:stiffness}, we could deduce that in the
$\tau\to0$ limit our formulae lead to the \Ito interpretation of
\eqref{eq:passive-tracer}.

We can also check on the simulations (done for small enough time-step
that we don't even need the sophistications of subsection
\ref{ssec:stiffness}) that this is the case.  Our simulation of
particle separation is equivalent to tracing a single particle in a
linear (in the spatial variable) velocity field, that is $\vec U =
\sigma \vec R$ recalling \eqref{eq:def-sigma}.  Hence the Lyapunov
exponent of pair separation of the passive tracer is described by
\eqref{eq:passive-tracer} with the identification $\vec R = \vec r$
and $\vec U = \vec v$.  Now the Lyapunov exponent obtained from
\eqref{eq:passive-tracer} depends on the convention we adopt for the
latter.  Taking the \Ito convention leads to substituting $\dd \vec r
= \dd S \vec r$ (recall \eqref{eq:def-dS}) into \eqref{eq:log-Ito} and
we get
$$
  \left\langle \frac{\dd \log r^2}{2 \dd t} \right\rangle
=
  -\frac{1}{r^4} C_{ik,jl} r_i r_k r_j r_l
  +\frac{1}{2 r^2} C_{ik,il} r_k r_l
=
  D_1 (d-4\wp) (d-1)
$$
In the simulations this is the value we get in the $\tau\to0$ limit.

In comparison, Stratonovich interpretation of
\eqref{eq:passive-tracer} would require considering $\dd r_i = \dd
S_{ij} r_j + (1/2) \langle \dd S_{ik} \dd S_{kl} \rangle r_l = \dd
S_{ij} r_j + (1/2) C_{ik,kl} r_l = \dd S_{ij} r_j + D_1 \wp (d+2)(d-1)
r_i$ which would give a Lyapunov exponent shifted by the non-zero $D_1
\wp (d+2)(d-1)$.

The fact that for inertial particles the Lagrangian equation
\eqref{eq:passive-tracer} must always be taken within the \Ito
convention is due to the fundamental time-irreversible behavior of
such particles.  Indeed, particles have some knowledge of fluid
velocity in the past, which they forget only after about a time of the
order of the Stokes time $\tau$, however they know nothing of the
future.  It is this smoothing of the fluid velocity at time-scale
$\tau$, using {\em only past events}, that intuitively explains why
the $\tau \to 0$ limit leads invariably to a passive tracer which
obeys \eqref{eq:passive-tracer} with \Ito convention.

\section{Stratonovich convention for linearized velocity?}

This is just a short note to explain why one cannot naively {\em
first} linearize \eqref{eq:passive-tracer} (in the position variable)
and {\em then} change reading convention, say from \Ito to
Stratonovich.  The two operations just do not commute with one
another, and the physically meaningful order is the other one:
linearizing the effective evolution of particles.

To put very simply what fails, note that when passing from \Ito to
Strato\-no\-vich interpretation in \eqref{eq:passive-tracer}, an
additional drift term of the form $(\vec U . \vec \nabla) \vec U$
appears: the Stratonovich interpretation of \eqref{eq:passive-tracer}
corresponds to the \Ito SDE
\begin{equation}
\label{eq:pass-trac-Str}
  \dd \vec R(t)
=
  \dd \vec U_t(\vec R(t))
 +\frac{1}{2}
  \langle (\dd\vec U_t . \vec \nabla) \dd\vec U_t \rangle \, (\vec R(t))
\end{equation}
Clearly, linearization of this term involves linearization of the {\em
derivative} of $\dd\vec U_t$, that comes from the quadratic part of
$\dd\vec U_t$, which would have been lost if we were to linearize
$\dd\vec U_t$ first and change convention after.

Coming back to the simpler notation $\vec U$, we can write the
gradient of $(\vec U . \vec\nabla) \vec U$, in coordinate notation:
$$
  \partial_k [ (U_j \partial_j) U_i]
=
  (\partial_k U_j) (\partial_j U_i) + U_j \partial_j \partial_k U_i
=
  \sigma_{ij} \sigma_{jk} + U_j \partial_k \sigma_{ij}
$$
If the velocity field $\vec U$ has spatially homogeneous statistics,
then $\langle (U_j \partial_j) U_i \rangle$ is independent of position
so its gradient is 0, implying that the Stratonovich term vanishes
when we linearize \eqref{eq:pass-trac-Str}; we also have the identity:
$$
  \langle \sigma_{ij} \sigma_{jk} \rangle
 +\langle U_j \partial_k \sigma_{ij} \rangle
=
  0
$$

Had we done linearization of the velocity field first, that is
supposing $\sigma$ to be constant, we would have $\partial_k
\sigma_{ij} \equiv 0$, and the linearization of the
Stra\-to\-no\-vich term would, {\em incorrectly}, reduce to the
non-zero quantity $\frac{1}{2} r_k \langle \sigma_{ij} \sigma_{jk}
\rangle$ proportional to (with contraction on repeated indices)
$\frac{1}{2} C_{ij,jk} r_k = 2 D_1 (d+2) (d-1) \wp r_i$, vanishing
only in the incompressible case $\wp = 0$.

\section{Positive even order moments of pair separation}
\label{app:pos-even-mom}

\subsection{Arbitrary dimensions}

Let us rewrite equation \eqref{eq:Anderson-multi} as two coupled
first-order equations for $\vec\psi$ and $\dot{\vec{\smash[t]\psi}} =
\dd\vec\psi/\dd t$ in coordinate notation:
$$
  \dd \psi_i = \dot\psi_i \dd t \,, \qquad
  \dd \dot\psi_i = -E \psi_i \dd t + \tau^{-1} (\dd S_{ij}) \psi_j
$$
(recall definition of $E$ from \eqref{def:E,V} and of $\dd S$ from
\eqref{eq:def-dS}).  The noteworthy fact is that this is a homogeneous
equation in $\vec\psi$, $\dot{\vec{\smash[t]\psi}}$, so that moments
of the same positive total order in the components will verify a
closed system of equations:
\begin{align*}
  \dd \langle \prod_{i,j=1}^d \psi_i^{m_i} \dot\psi_j^{n_j} \rangle
=\!&\phantom{+}
  \sum_k \langle
    m_k \psi_k^{m_k-1} (\dd \psi_k)
    \prod_{i \neq k} \psi_i^{m_i} \prod_j \dot\psi_j^{n_j}
  \rangle
\\
&+
  \sum_k \langle
    n_k \dot\psi_k^{n_k-1} (\dd \dot\psi_k)
    \prod_i \psi_i^{m_i} \prod_{j \neq k} \dot\psi_j^{n_j}
  \rangle
\\
&+
  \sum_k \langle
    {\textstyle \frac{n_k (n_k-1)}{2}}
    \dot\psi_k^{n_k-2} (\dd \dot\psi_k) (\dd \dot\psi_k)
    \prod_i \psi_i^{m_i} \prod_{j \neq k} \dot\psi_j^{n_j}
  \rangle
\\
&+
  \sum_{k \neq l} \langle
    n_k n_l \dot\psi_k^{n_k-1} \dot\psi_l^{n_l-1}
      (\dd \dot\psi_k) (\dd \dot\psi_l)
    \prod_i \psi_i^{m_i} \prod_{j \neq k,l} \dot\psi_j^{n_j}
  \rangle
\intertext{which, upon replacing the $\dd\psi$ and the $\dd\dot\psi$
  by their expressions, leads to}
=&\phantom{+}
  \sum_k m_k \langle
    \psi_k^{m_k-1} \dot\psi_k
    \prod_{i \neq k} \psi_i^{m_i} \prod_j \dot\psi_j^{n_j}
  \rangle \,\dd t
\\
&-
  \sum_k E n_k \langle
    \dot\psi_k^{n_k-1} \psi_k
    \prod_i \psi_i^{m_i} \prod_{j \neq k} \dot\psi_j^{n_j}
  \rangle \,\dd t
\\
&+
  \sum_k \sum_{p,q}
    {\textstyle \frac{n_k (n_k-1)}{2}} \tau^{-2} C_{kp,kq} \langle
      \dot\psi_k^{n_k-2} \psi_p \psi_q
      \prod_i \psi_i^{m_i} \prod_{j \neq k} \dot\psi_j^{n_j}
    \rangle \,\dd t
\\
&+
  \sum_{k \neq l} \sum_{p,q} n_k n_l \tau^{-2} C_{kp,lq} \langle
    \dot\psi_k^{n_k-1} \dot\psi_l^{n_l-1}
      \psi_p \psi_q
    \prod_i \psi_i^{m_i} \prod_{j \neq k,l} \dot\psi_j^{n_j}
  \rangle \,\dd t
\end{align*}

Introduce the multi-coefficients $\ul{m}$ and $\ul{n}$ for the $m_i$
and $n_j$, and use the notation $\ul{m}^k$ for adding 1 to the
$k^\text{th}$ coefficient of $\ul{m}$, and $\ul{m}_k$ for substracting
1 from the $k^\text{th}$ coefficient of $\ul{m}$, etc.  Also denote
$c_{\ul{m},\ul{n}} = \langle \prod_{i,j=1}^d \psi_i^{m_i}
\dot\psi_j^{n_j} \rangle$.  Then we have the closed system for the
$c_{\ul m, \ul n}$ with $[\ul m]+[\ul n]$ constant (brackets stand for
the size of the multi-coefficient):
\begin{multline}
\label{eq:even-moments}
  \frac{\dd c_{\ul m, \ul n}}{\dd t}
=
   \sum_k m_k  c_{\ul m_k, \ul n^k}
  -\sum_k E n_k c_{\ul m^k, \ul n_k}
\\
  +\sum_k \sum_{p,q}
     {\textstyle \frac{n_k (n_k-1)}{2}} \tau^{-2} C_{kp,kq}
     c_{\ul m^{p,q}, \ul n_{k,k}}
  +\sum_{k \neq l} \sum_{p,q} n_k n_l \tau^{-2} C_{kp,lq}
     c_{\ul m^{p,q}, \ul n_{k,l}}
\end{multline}

The idea which permits to compute the growth rate of $|\vec\psi|^{2N}$
for $N \in \NN$ is that it should be given by the topmost eigenvalue
(Lyapunov exponent) of \eqref{eq:even-moments} considered as a system
of linear differential equations on $c_{\ul m, \ul n}$ for $\sum m_i +
\sum n_i = 2N$.  The system in question is linear (in the $c_{\ul m,
\ul n}$) with constant coefficients (i.e.\ independent of $t$), so it
has an associated matrix, and what we need is the top eigenvalue of
this matrix.  In general this eigenvalue should be the exponential
growth rate of $|\vec\psi|^{2N}$.

\subsection{Simplified version in 2D}

In the two-dimensional case a somewhat simpler formulation of the
above may be given.  Let us introduce $\psi = e^{t/2\tau} r$,
$\dot\psi = \dd\psi/\dd t$ and their conjugates $\bar\psi$ and
$\dot{\bar{\smash[t]\psi}}$.  Recall \eqref{eq:Anderson-complex}:
$$
  -\frac{\dd^2\psi}{\dd t^2} + V \psi
=
  E \psi
$$
Furthermore, denoting
$$
  c^n_{k,l}
=
  \langle
    \psi^k \dot\psi^{n-k} \bar\psi^l \dot{\bar{\smash[t]\psi}}^{n-l}
  \rangle
$$
the equivalent of the system \eqref{eq:even-moments} is
\begin{align*}
  \frac{\dd c^n_{k,l}}{\dd t}
=&\phantom{+}
    k c^n_{k-1,l} - E(n-k) c^n_{k+1,l}
  + l c^n_{k,l-1} - E(n-l) c^n_{k,l+1}
\\
& + {\textstyle \frac{(n-k)(n-k-1)}{2}} (\beta_L-\beta_N) c^n_{k+2,l}
  + {\textstyle \frac{(n-l)(n-l-1)}{2}} (\beta_L-\beta_N) c^n_{k,l+2}
\\
& + (n-k)(n-l)(\beta_L+\beta_N) c^n_{k+1,l+1}
\end{align*}
This system has been derived in \cite{MR1953940} for similar purposes
as here, in the framework of the Anderson localization problem.

\bibliographystyle{plain}
\bibliography{2dlyap}

\begin{thebibliography}{10}

\bibitem{MR1225604}
Milton Abramowitz and Irene~A. Stegun, editors.
\newblock {\em Handbook of mathematical functions with formulas, graphs, and
  mathematical tables}.
\newblock Dover Publications Inc., New York, 1992.
\newblock Reprint of the 1972 edition.

\bibitem{Collision}
J.~Bec, A.~Celani, M.~Cencini, and S.~Musacchio.
\newblock Clustering and collisions of heavy particles in random smooth flows.
\newblock Preprint at \texttt{arXiv:nlin.CD/0407013}.

\bibitem{rainColl}
G.~Falkovich, A.~Fouxon, and M.~G. Stepanov.
\newblock Acceleration of rain initiation by cloud turbulence.
\newblock {\em Nature}, 419:151--4, 2002.

\bibitem{MR1878800}
G.~Falkovich, K.~Gaw{\c{e}}dzki, and M.~Vergassola.
\newblock Particles and fields in fluid turbulence.
\newblock {\em Rev. Modern Phys.}, 73(4):913--975, 2001.

\bibitem{2dlyapShort}
A.~Fouxon and P.~Horvai.
\newblock Inertial particles and the {A}nderson localization problem.
\newblock {\em In consideration for Phys. Rev. Lett.}, 2006.
\newblock Preprint will be in \texttt{arXiv:cond-mat}.

\bibitem{MR0187859}
Bertrand~I. Halperin.
\newblock Green's functions for a particle in a one-dimensional random
  potential.
\newblock {\em Phys. Rev. (2)}, 139:A104--A117, 1965.

\bibitem{MaxRi}
M.~R. Maxey and Riley J.
\newblock Equation of motion of a small rigid sphere in a nonuniform flow.
\newblock {\em Phys. Fluids}, 26:883--889, 1983.

\bibitem{InAgg}
B.~Mehlig, M.~Wilkinson, K.~Duncan, T.~Weber, and Ljunggren M.
\newblock On the aggregation of inertial particles in random flows.
\newblock {\em Submitted to Phys. Rev. E (3)}, 2005.
\newblock Preprint at \texttt{arXiv:cond-mat/0410518}.

\bibitem{PRL2506021}
Bernhard Mehlig and Michael Wilkinson.
\newblock Coagualtion by random velocity fields as a {K}ramers problem.
\newblock {\em Phys. Rev. Lett.}, 92(25):250602--1--4, 2004.

\bibitem{MR1897722}
Leonid~I. Piterbarg.
\newblock The top {L}yapunov exponent for a stochastic flow modeling the upper
  ocean turbulence.
\newblock {\em SIAM J. Appl. Math.}, 62(3):777--800 (electronic), 2001/02.

\bibitem{MR1953940}
H.~Schomerus and M.~Titov.
\newblock Statistics of finite-time {L}yapunov exponents in a random
  time-dependent potential.
\newblock {\em Phys. Rev. E (3)}, 66(6):066207, 11, 2002.

\end{thebibliography}

\newpage

\begin{figure}
\centerline{\includegraphics[width=.65\textwidth]{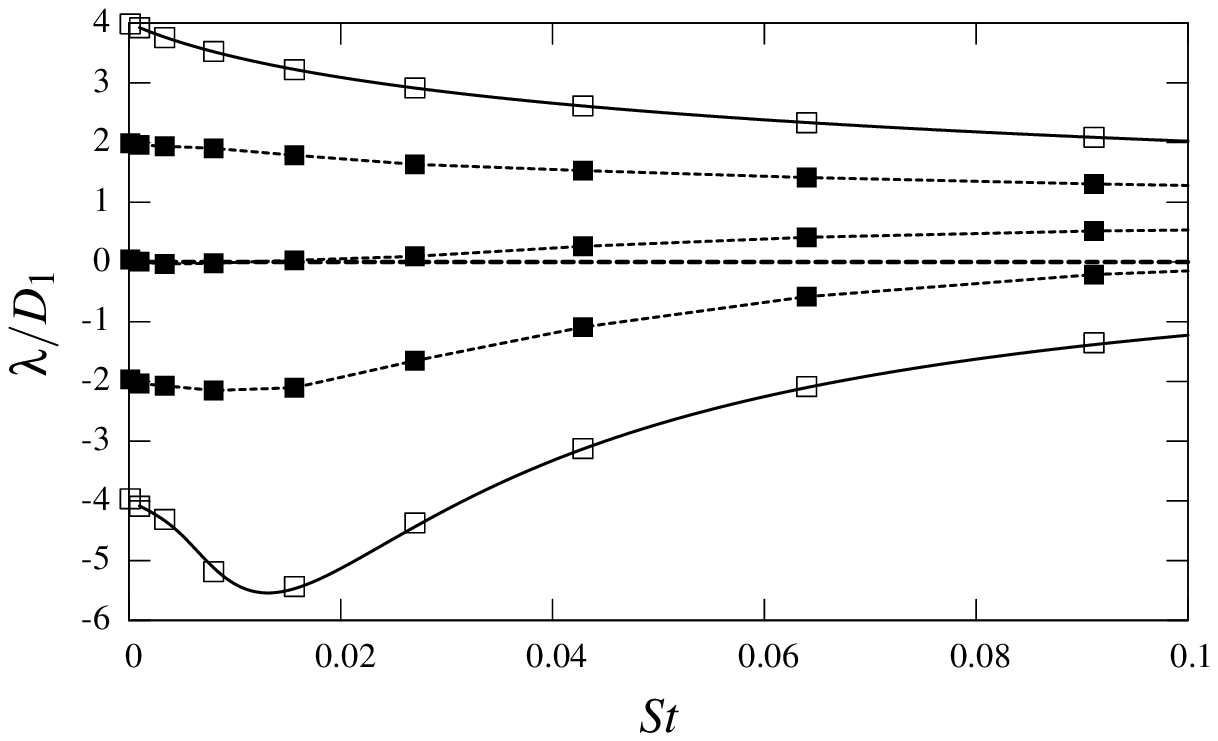}}

\centerline{\includegraphics[width=.65\textwidth]{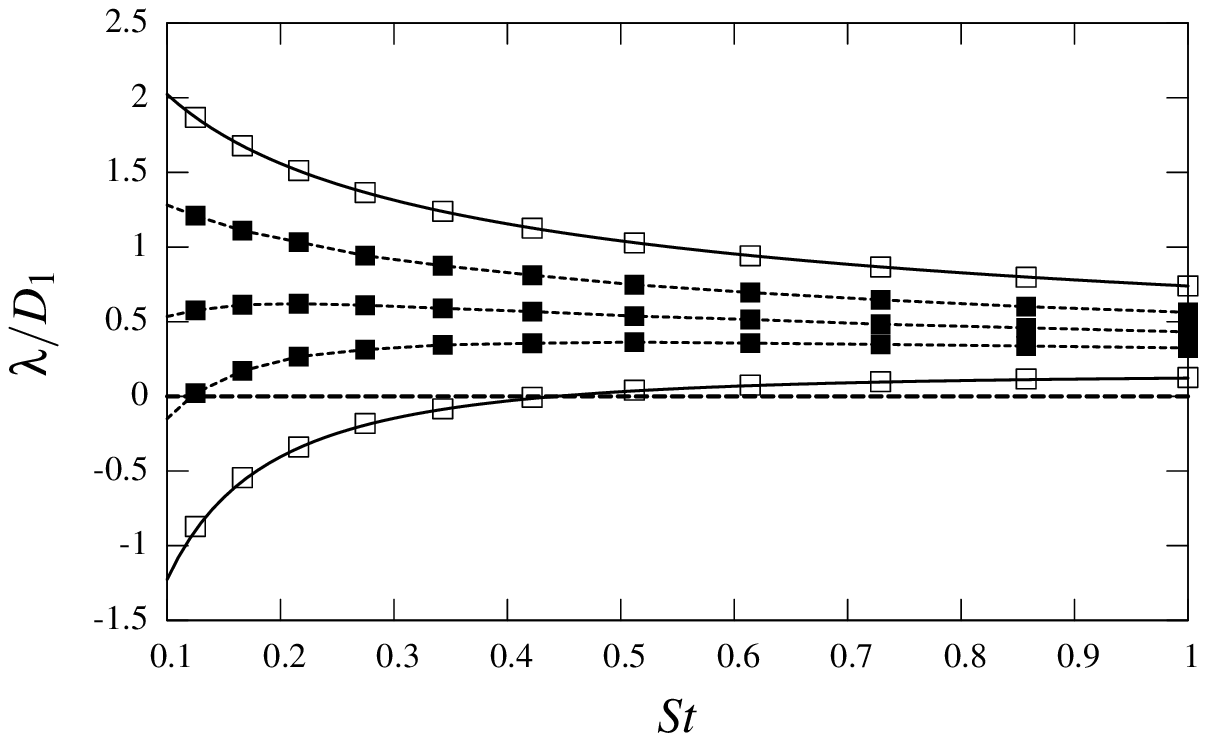}}

\centerline{\includegraphics[width=.65\textwidth]{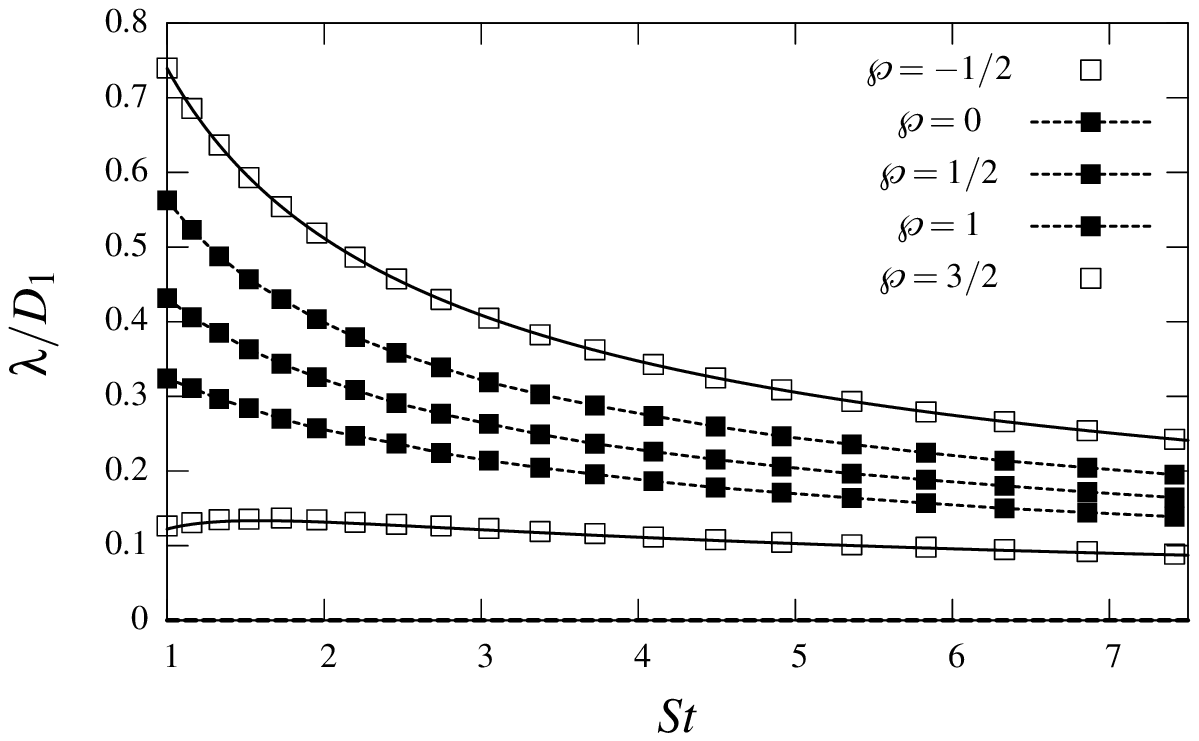}}
\caption{Adimensionalized Lyapunov exponent in function of Stokes
  number for different values of $\wp$.  We see perfect agreement
  between theory (solid lines only, i.e.\ those with the empty boxes)
  and numerics (boxes, solid or empty) for $\wp=-1/2$ and $\wp=3/2$.}
\label{fig:Lyap-plots}
\end{figure}

\newpage

\begin{figure}
\centerline{\includegraphics[width=.8\textwidth]{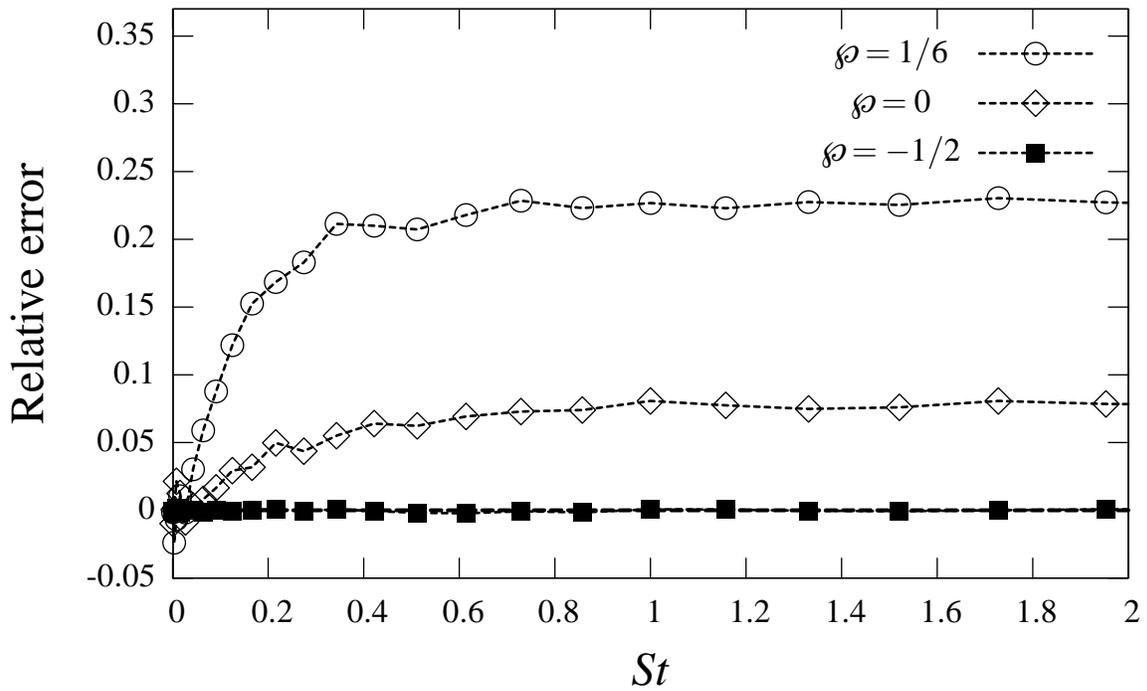}}
\caption{Relative error of numerical results with respect to the
  formula conjectured in \cite{MR1897722} and reproduced in our paper
  as formula \eqref{eq:conj-Piterbarg}, for $\wp < 1/2$.  A zoom (not
  represented) on the curves for very small $\mathit{St}$ is not
  incompatible with the prediction (according to the divergent
  asymptotic expansion in \cite{PRL2506021}) that all derivatives of
  the relative error curves should vanish at $\mathit{St}=0$, but
  quality of our data in that range is too poor.}
\end{figure}

\begin{figure}
\centerline{\includegraphics[width=.8\textwidth]{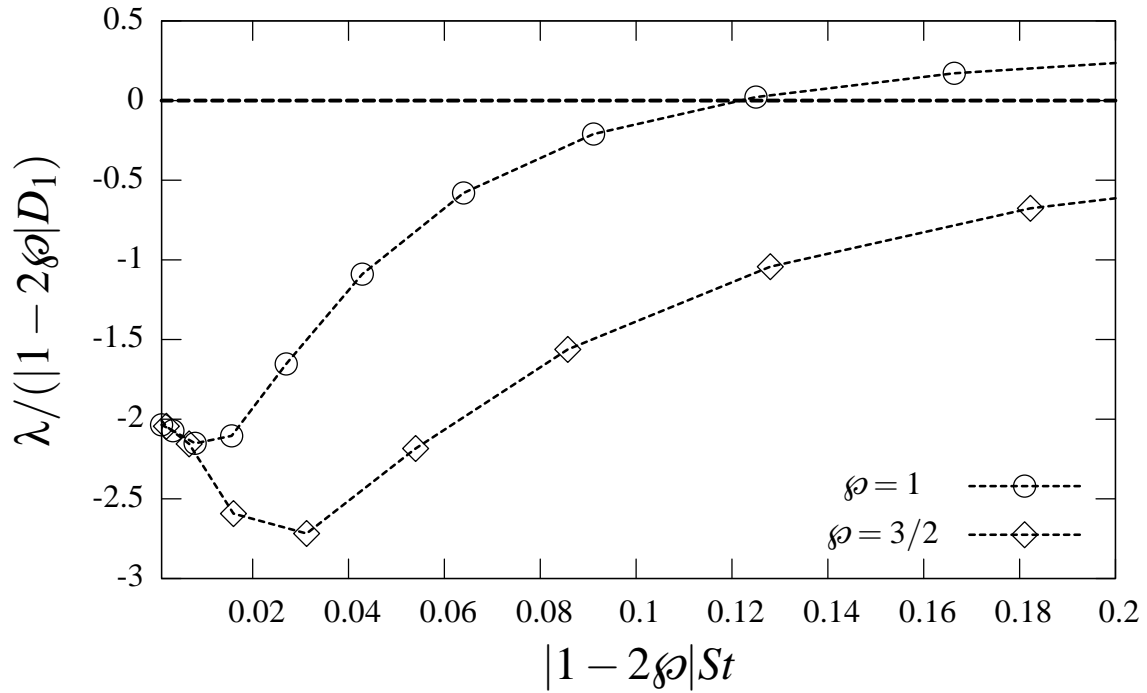}}
\caption{Non-collapse of numeric curves for $\wp>1/2$, contradicting
  \eqref{eq:conj-scaling}} 
\end{figure}

\newpage

\begin{figure}
\centerline{\includegraphics[width=.65\textwidth]{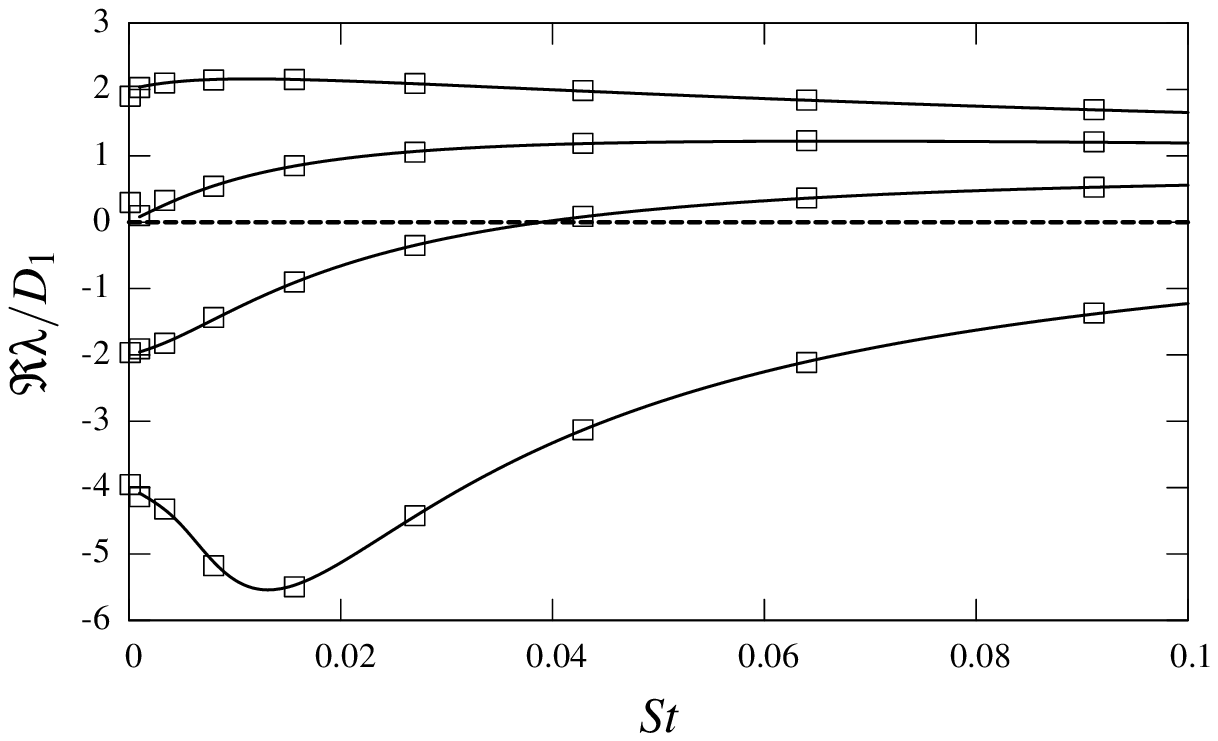}}

\centerline{\includegraphics[width=.65\textwidth]{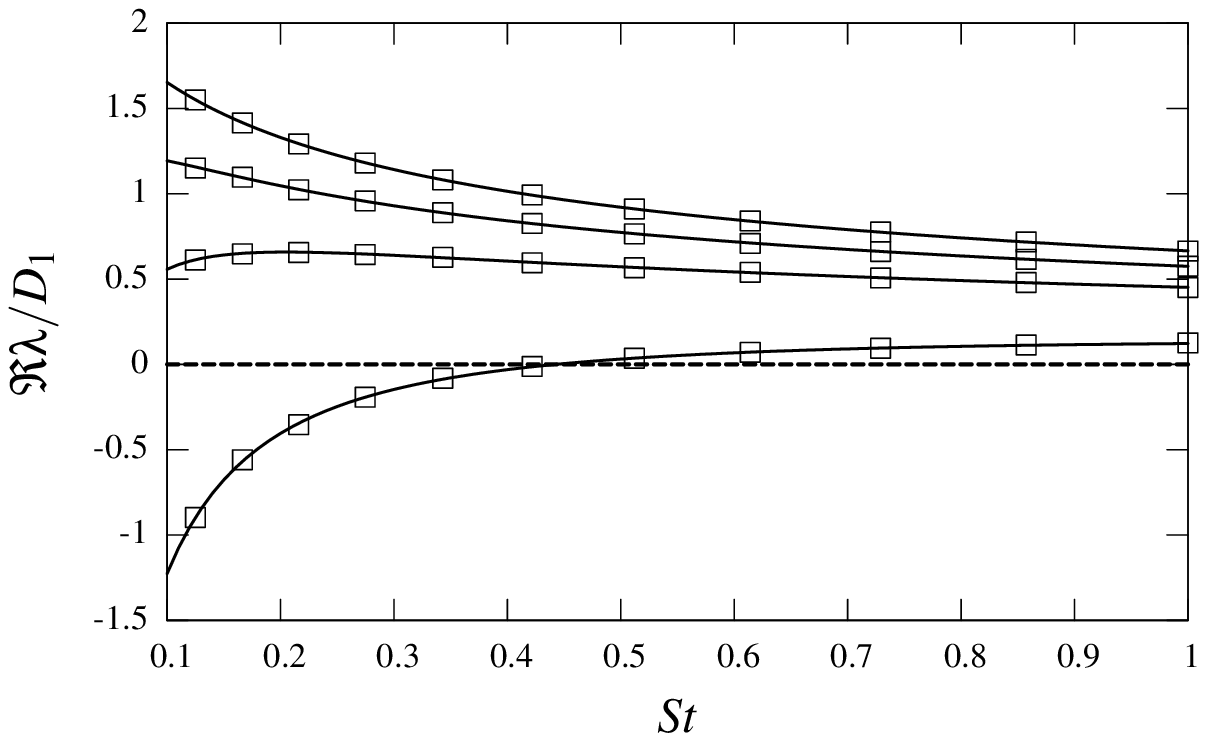}}

\centerline{\includegraphics[width=.65\textwidth]{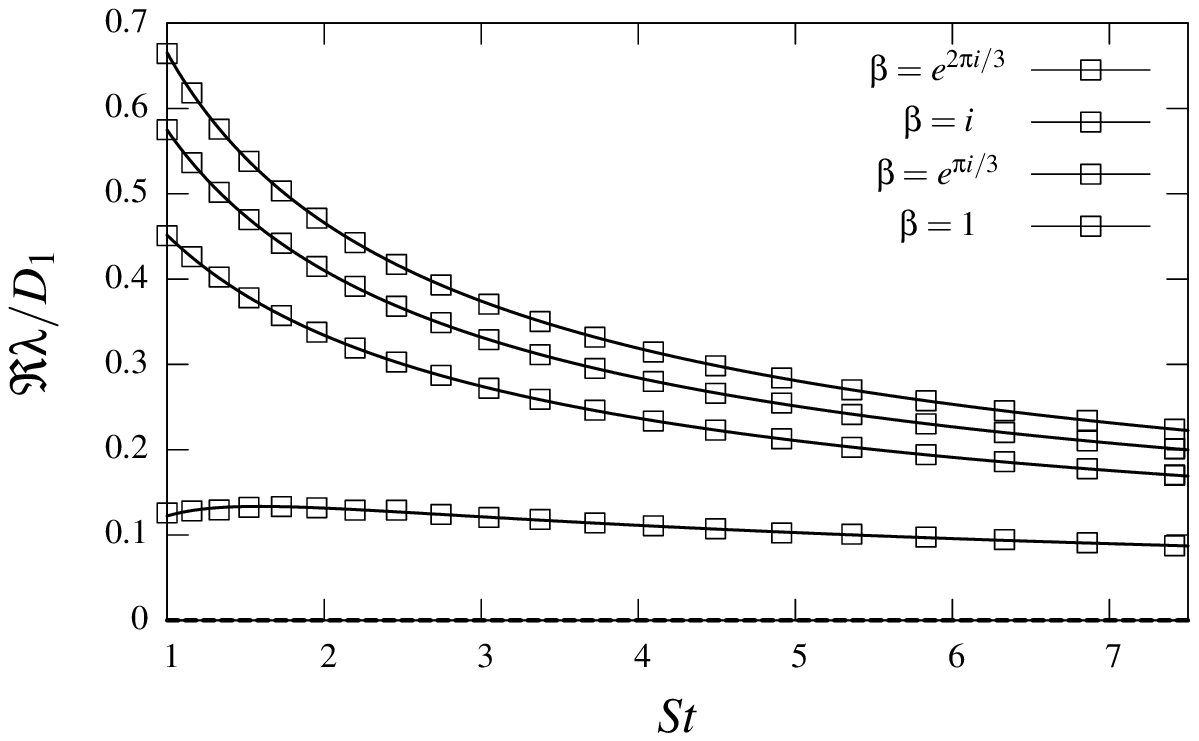}}
\caption{Real part of adimensionalized Lyapunov exponent in function of
  Stokes number for different values of $\beta$.  We see perfect agreement
  between theory (solid lines) and numerics (empty boxes)}
\label{fig:Lyap-L-plots}
\end{figure}

\newpage

\begin{figure}
\centerline{\includegraphics[width=.65\textwidth]{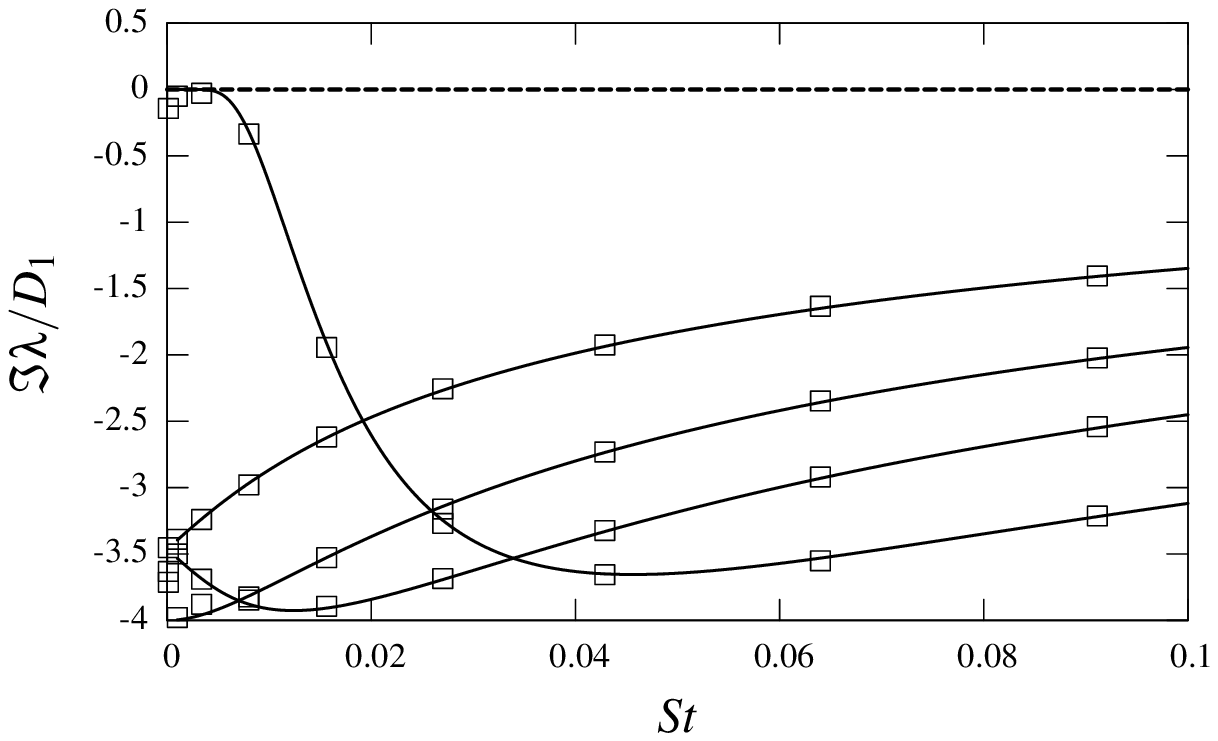}}

\centerline{\includegraphics[width=.65\textwidth]{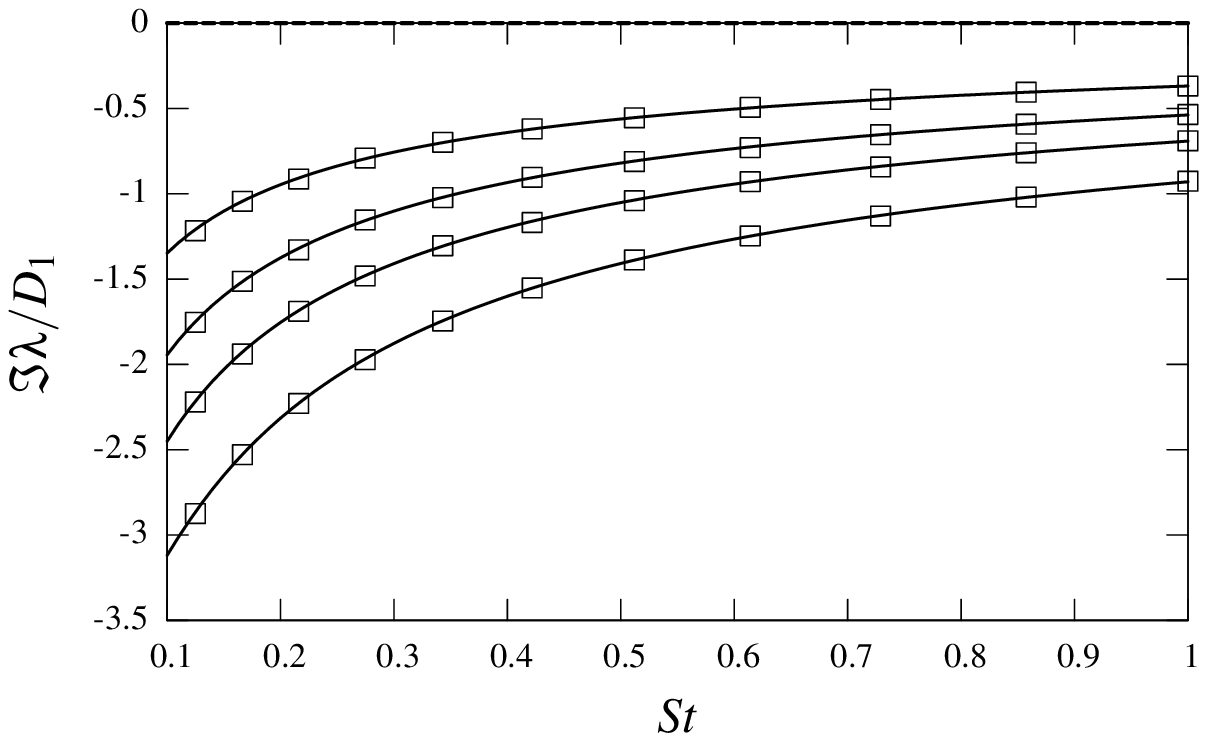}}

\centerline{\includegraphics[width=.65\textwidth]{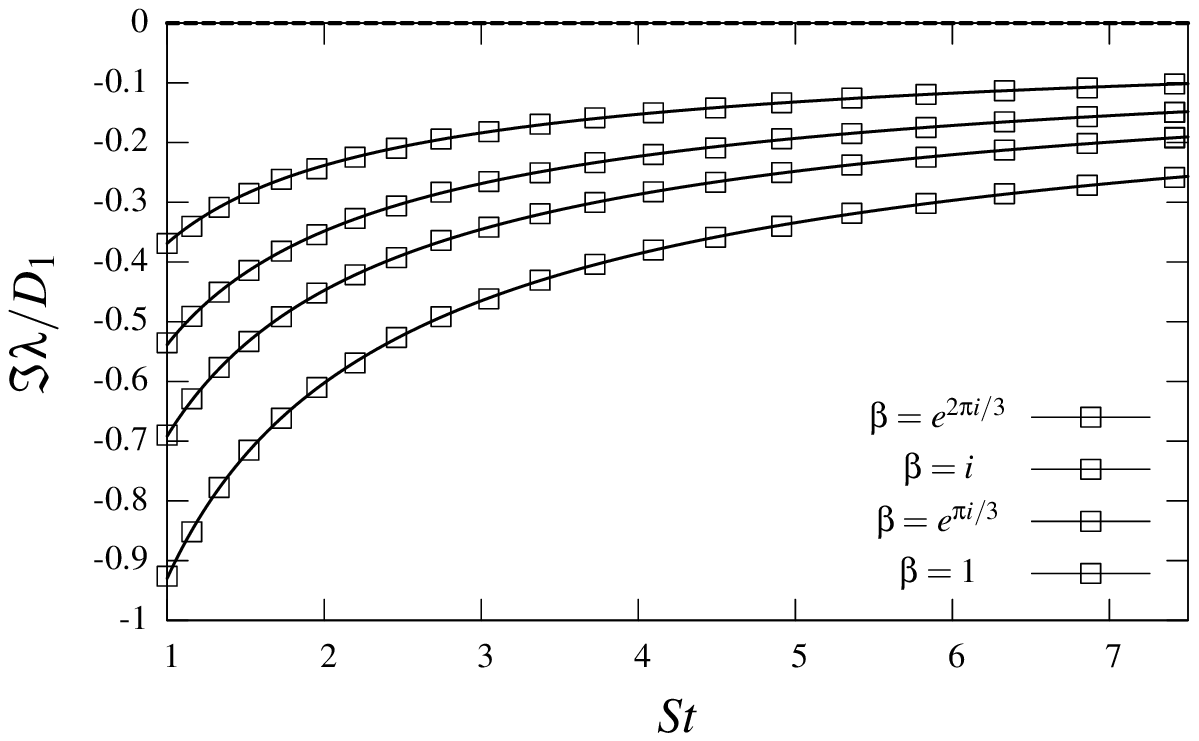}}
\caption{Imaginary part of adimensionalized Lyapunov exponent in function of
  Stokes number for different values of $\beta$.  We see perfect agreement
  between theory (solid lines) and numerics (empty boxes)}
\label{fig:Lyap-Li-plots}
\end{figure}

\end{document}